\documentclass[aps,prb,preprint,amsmath,amssymb,floatfix,nofootinbib]{revtex4-2}
\usepackage{graphicx}
\usepackage{xcolor}
\usepackage{bm}
\begin{document}
\title{Time-dependent resonating Hartree--Fock theory for correlated electron--lattice dynamics: a nonorthogonal multiconfiguration variational approach}
\author{Norikazu Tomita}
\email{tomita@sci.kj.yamagata-u.ac.jp}
\affiliation{Department of Physics, Faculty of Science, Yamagata University, 1-4-12 Kojirakawa-machi, Yamagata, Yamagata 990-8560, Japan}
\author{Akira Takahashi}
\affiliation{Graduate School of Engineering, Nagoya Institute of Technology, Showa-ku, Gokiso-cho, Nagoya, Aichi 466-8555, Japan}
\date{\today}
\begin{abstract}
We develop a tractable time-dependent theory, time-dependent resonating Hartree--Fock (TD-ResHF), in which both the electrons and the lattice are treated quantum mechanically. The wave function is a superposition of a small number of nonorthogonal configurations, each the product of a Slater determinant and a lattice coherent state, and all its parameters evolve according to the McLachlan variational principle. We apply the method to the extended Hubbard model with Su--Schrieffer--Heeger-type electron--lattice coupling, in which the nearest-neighbor interaction is raised linearly in time. With five configurations, it reproduces the exact solutions for chains of 6 and 10 sites, including the correlation functions, the lattice distortion, and the energy, to within 1--2\%. The calculations show that the nonadiabatic and adiabatic regimes of the electrons are distinguished by the excess energy and by the oscillation of the correlation functions after the ramp, whereas the lattice stays close to its initial distortion even for the slowest ramp and leaves the correlation functions displaced from their static values at the final interaction. For 14 sites, a size too large for exact diagonalization, the post-ramp oscillation, which is caused by nonadiabatic excitation of the electrons, is visualized in the wave function itself as a transfer of weight between configurations whose shapes hardly change. To keep the time evolution stable, the configuration set is reconstructed when the configurations become nearly linearly dependent.
\end{abstract}
\maketitle
\clearpage
% Introduction (clean English version, 2026-11-02). Citation keys are placeholders:
% Simenel, CurchodMartinez, YonemitsuNasu, Aoki, Yonemitsu, Koshihara, Rajpurohit, LuLiHuang, Miyashita, OkamotoMirmohammadi,
% WhiteFeiguin, Daley, Haegeman, Matsueda, Shi, Wang, Ido, Zhao, BorrelliGelin,
% JoubertDoriol, Werther, Asaad, Hojlund, Schrader, Fukutome, Shida, Watanabe

\section{Introduction}\label{sec:intro}

Nonequilibrium dynamics of quantum many-body systems is a research theme across many areas of physics, including nuclear dynamics \cite{Simenel}, molecular processes \cite{CurchodMartinez}, and photoinduced phenomena in solids \cite{YonemitsuNasu,Aoki}. What makes the problem attractive is that a driven system does not simply move from one equilibrium state to another: it is left in a superposition of many excited states, and its subsequent motion contains coherent oscillations and states that are not found in equilibrium. What makes it difficult is that the degrees of freedom involved evolve on different time scales, so that which of them follow the drive and which lag behind cannot be read off from equilibrium properties, while the dimension of the wave function grows exponentially with the system size. Understanding such dynamics is also the basis for controlling material properties, as in ultrafast switching.

In condensed-matter physics, the coupled dynamics of correlated electrons and the lattice gives rise to rich nonequilibrium phenomena, of which photoinduced phase transitions in organic charge-transfer solids are prominent examples \cite{YonemitsuNasu,Yonemitsu}. Experiments have found hidden states that are unique to the photoexcited state, and have shown that in materials with strong electron--lattice coupling, such a state is formed within tens of femtoseconds, accompanied by coherent lattice vibrations \cite{Koshihara}. A theoretical description of such a process requires a time evolution of the coupled system in which electronic many-body correlations and the quantum nature of the lattice motion, including zero-point fluctuations and quantum coherence between different lattice distortions, are both taken into account. What is needed is a method that does so at a tractable computational cost.

Some methods follow the electronic correlations without a lattice. The time-dependent variational Monte Carlo method uses the McLachlan variational principle with many-variable correlated wave functions, but it has been benchmarked only on Hubbard models, which have no lattice degrees of freedom, for continuous times of the order of $10\hbar/t_0$, where $t_0$ is the hopping amplitude \cite{Ido}. When the lattice is included, it has mostly been treated classically. In first-principles real-time time-dependent density functional theory \cite{Rajpurohit,LuLiHuang} the ions are treated classically, and the lattice has also been treated classically in model calculations, in which the electrons are described by the time-dependent Schr\"odinger equation of small clusters or by mean-field theory \cite{Miyashita,OkamotoMirmohammadi,Yonemitsu}. A classical lattice, however, cannot describe the quantum nature of the lattice motion. The lattice can be treated quantum mechanically in several ways, but each has its limitations. Time-dependent density-matrix renormalization group (DMRG) and matrix-product-state variants of the time-dependent variational principle are powerful in one dimension \cite{WhiteFeiguin,Daley,Haegeman}, but for lattice vibrations they require a truncated local phonon basis, which limits the number of phonons that can be kept; a dynamical optimization of the local basis can reduce this cost in some cases \cite{Brockt}. Infinite-system variants such as the infinite time-evolving block decimation (iTEBD) \cite{Vidal} have also been applied to electron--phonon systems, again with the local phonon space truncated to a finite number of states per site (up to 12 in Ref.~\cite{Sous}). For the extended Hubbard--Holstein chain, the relaxation of photocarriers has been followed up to a time of about $100\hbar/t_0$ for weak electron--lattice coupling and chains of up to 12 sites \cite{Matsueda}. Variational treatments of non-Gaussian phonon states have been applied to the Holstein and Hubbard--Holstein models \cite{Shi,Wang}. Nonequilibrium dynamical mean-field theory (DMFT) has been extended to electron--phonon systems such as the Holstein--Hubbard model \cite{Aoki,WernerEckstein}, but it captures only local correlations in infinite dimensions. Superpositions of coherent states describe quantum vibrations well \cite{Zhao,BorrelliGelin}, but have mostly been applied to a single electron. Each of these methods is thus restricted in the dimensionality, the strength of the electron--lattice coupling, the range of the correlations, or the number of electrons, and a complementary approach is desirable.

One approach that can combine correlated electronic states with quantum lattice motion is the expansion in nonorthogonal basis functions whose parameters evolve in time. It has been developed extensively for molecular dynamics, in the form of multiple spawning, variational multiconfigurational Gaussians, and multiconfigurational Ehrenfest methods \cite{CurchodMartinez}. Time-dependent multiconfiguration methods have also been applied to the coupled electron--nuclear dynamics of laser-driven molecules \cite{Lotstedt}. In nuclear physics, a superposition of time-dependent mean-field trajectories has been proposed for collisions and tested on a model of pair transfer between two superfluids \cite{Verriere}, where the mean-field states follow their own equations of motion and the weights are determined variationally. These expansions share a numerical difficulty: when the basis functions approach each other, the overlap matrix becomes nearly singular, and the equations of motion become unstable. Proposed remedies include discarding or regularizing the singular directions \cite{JoubertDoriol}, reducing the number or the freedom of the basis functions \cite{Werther}, modifying the equations of motion to restore energy conservation \cite{Asaad} or to stabilize the evolution \cite{Hojlund}, and reoptimizing the basis at every time step \cite{Schrader}. In these studies, however, the nonorthogonal functions are Gaussians or determinants with fixed orbitals, applied to molecules or few-level models. To our knowledge, a time evolution in which the Slater determinants, the lattice coherent states, and the coefficients all evolve variationally has not been formulated for strongly correlated electron--lattice systems.

In this paper we develop a time-dependent extension of the resonating Hartree--Fock (ResHF) theory \cite{Fukutome}, which we call TD-ResHF. The ResHF wave function is a superposition of nonorthogonal Slater determinants, and we extend it to electron--lattice systems by attaching a coherent state of the lattice to each determinant \cite{Shida,Watanabe}. Each configuration is thus the product of a Slater determinant and a coherent state, and the lattice is treated in the full Fock space without a cutoff of the phonon number; the accuracy is controlled by the number of configurations, and each configuration can be inspected explicitly. The wave function is projected onto the symmetries of the Hamiltonian (translation, reflection, and spin flip), and the determinants, the coherent states, and the superposition coefficients all evolve according to the McLachlan variational principle. When the configurations become nearly linearly dependent and the overlap matrix becomes ill-conditioned, we restart the time evolution from a new set of configurations that represents the same state. This restart serves only to stabilize the calculation and does not modify the equations of motion; the details are given in Sec.~II and Appendix~A.

We apply the method to the extended Hubbard model coupled to Su--Schrieffer--Heeger (SSH)-type phonons, in which the nearest-neighbor interaction $V$ is increased linearly in time over a ramp time $\tau$ (in units of $\hbar/t_0$), as in the studies of linear ramps of the interaction in the Hubbard model without a lattice \cite{Sandri}. We first test TD-ResHF against exact time evolution in three cases: the electrons alone ($N=10$, $\tau=5$, 13, and 30), the coupled electron--lattice system with $N=6$ ($\tau=5$ and 13), where all phonon modes are included in the exact calculation, and the coupled system with $N=10$ ($\tau=5$, 13, and 100), where the exact calculation is restricted to the $\pi$ phonon mode. In all these cases, five configurations reproduce the exact results to within 1--2\%, including the oscillation after the ramp. We then apply the method to the coupled system with $N=14$ ($\tau=5$), a size beyond exact diagonalization.

The calculations lead to two main results. First, the difference between the electronic and the lattice time scales is made visible in a single time evolution. That the lattice is slower than the electrons is expected, but here its consequences can be followed quantitatively: the lattice distortion remains at its initial value for $\tau=5$ and $13$ and decreases only partly for $\tau=100$, where the ramp time is comparable to the phonon period, so that the correlation functions stay away from their static values at the instantaneous $V(t)$. For the electrons alone, in contrast, the correlation functions approach the static values as $\tau$ increases, and the difference between the two cases is the effect of the lattice. Second, the configurations and their weights give a picture of the oscillation after the ramp. At $N=14$, the oscillation is described by a transfer of weight between two groups of configurations, while the configurations themselves change little: one group has a strong charge-density wave, and the other has a weak charge-density wave with a local spin density.

TD-ResHF thus treats correlated electrons and quantum lattice motion together with a small number of configurations, and the configurations provide a physical picture of the dynamics. The paper is organized as follows. Section~\ref{sec:method} describes the model and the method. Section~\ref{sec:results} presents the comparison with the exact solutions, the dependence on the ramp time and on the number of configurations, and the results for $N=14$. Section~\ref{sec:discussion} summarizes the results, discusses the physical picture that emerges, and comments on the method and on open questions. The reconstruction of the configuration set and the exact-diagonalization calculations are described in the appendices.

\section{Method and model}\label{sec:method}

\subsection{Model}\label{sec:model}

We consider the extended Hubbard model coupled to SSH-type phonons on a ring of $N$ sites,
\begin{align}
H &= -t_0\sum_{\ell\sigma}\bigl[1+\alpha(q_\ell-q_{\ell+1})\bigr]\bigl(a^\dagger_{\ell\sigma}a_{\ell+1\sigma}+\mathrm{h.c.}\bigr)
+U\sum_\ell n_{\ell\uparrow}n_{\ell\downarrow}+V\sum_\ell n_\ell n_{\ell+1} \nonumber\\
&\quad+\sum_\ell\Bigl[\frac{p_\ell^2}{2M}+\frac K2\,(q_{\ell+1}-q_\ell)^2\Bigr],
\label{eq:H}
\end{align}
where $t_0$ is the hopping amplitude, $n_\ell=n_{\ell\uparrow}+n_{\ell\downarrow}$, and $q_\ell$ and $p_\ell$ are the displacement and the momentum of the lattice at site $\ell$. Throughout this paper the system is at half filling, with $N/2$ electrons of each spin ($N_\uparrow=N_\downarrow=N/2$, $N$ even) and periodic boundary conditions. The lattice is expanded in normal modes,
\begin{equation}
q_\ell=\sum_k\sqrt{\frac{\hbar}{2MN\omega(k)}}\,\bigl(b_k+b^\dagger_{-k}\bigr)e^{ik\ell},\qquad
\omega(k)=2\sqrt{K/M}\,\bigl|\sin(k/2)\bigr| ,
\end{equation}
and we write $\omega\equiv\omega(\pi)$ for the frequency of the zone-boundary ($\pi$) mode. The uniform mode $k=0$ is a rigid translation of the lattice, which changes neither the bond lengths $q_\ell-q_{\ell+1}$ nor the lattice energy; it is decoupled from the electrons and is dropped. We use units in which $\hbar=t_0=1$, so that energies are measured in units of $t_0$ and time in units of $\hbar/t_0$. The parameters are $U=4$, $K=0.25$, $\alpha=0.29$, and $\omega=0.066$, and the interaction $V$ is a function of time, $V=V(t)$, which is specified in Sec.~\ref{sec:results}.

The observables are the charge, spin, and bond structure factors at wave number $\pi$,
\begin{align}
N(\pi)&=\frac1N\Bigl\langle\Bigl(\sum_\ell(-1)^\ell n_\ell\Bigr)^2\Bigr\rangle,\qquad
S(\pi)=\frac1N\Bigl\langle\Bigl(\sum_\ell(-1)^\ell S^z_\ell\Bigr)^2\Bigr\rangle,\\
B(\pi)&=\frac1N\Bigl\langle\Bigl(\sum_\ell(-1)^\ell B_\ell\Bigr)^\dagger\Bigl(\sum_\ell(-1)^\ell B_\ell\Bigr)\Bigr\rangle,
\end{align}
with $S^z_\ell=\frac12(n_{\ell\uparrow}-n_{\ell\downarrow})$ and $B_\ell=\sum_\sigma(c^\dagger_{\ell\sigma}c_{\ell+1\sigma}+\mathrm{h.c.})$, and the dimerization amplitude of the lattice,
\begin{equation}
D=2\Bigl[\langle Q_\pi^2\rangle-\frac{1}{2MN\omega}\Bigr]^{1/2},\qquad Q_\pi=\frac1N\sum_\ell(-1)^\ell q_\ell .
\end{equation}
Here $Q_\pi$ is the staggered component of the lattice displacement and the second term in the bracket is the zero-point contribution, so that $D$ is the amplitude of the alternation of the bond lengths $q_\ell-q_{\ell+1}$.

\subsection{Wave function}

The wave function is a superposition of nonorthogonal configurations, each of which is the product of a Slater determinant and a coherent state of the lattice, projected onto the symmetries of the Hamiltonian,
\begin{equation}
|\Phi(t)\rangle=\sum_{m=1}^{M_c}C_m(t)\,\hat P\,\bigl[\,|U^\uparrow_m(t)\rangle|U^\downarrow_m(t)\rangle|z_m(t)\rangle\bigr].
\label{eq:Phi}
\end{equation}
Here $U^\sigma_m$ is an $N\times N_\sigma$ matrix of occupied orbitals (the orbitals of up and down spins are different), $N_\sigma=N/2$, and
\begin{equation}
|z_m\rangle=\exp\Bigl(-\frac12\sum_k|z_{m,k}|^2\Bigr)\exp\Bigl(\sum_k z_{m,k}b^\dagger_k\Bigr)|0\rangle
\end{equation}
is the coherent state with the complex amplitudes $z_{m,k}$. $M_c$ is the number of configurations (NBM in the figures), and $\hat P$ projects onto translation, reflection, and spin flip (so-called half-projection). Under a translation by $n$ sites the coherent-state amplitudes transform as $z_k\to e^{-ikn}z_k$, and under a reflection as $z_k\to-z_{-k}$, while the rows (site indices) of the orbital matrices $U^\sigma_m$ are shifted cyclically by $n$ or reversed, respectively. The projection makes the sum over the elements of the symmetry group in all matrix elements, but it does not change the structure of the equations, and we write them below without it.

The matrix elements between two configurations are given by closed expressions. For the Slater determinants, $\langle m|n\rangle=\prod_\sigma\det\mathcal O^\sigma_{mn}$ with the overlap matrix of the orbitals $\mathcal O^\sigma_{mn}=U^{\sigma\dagger}_mU^\sigma_n$, and the interstate density matrix $W^\sigma_{mn}=U^\sigma_n\bigl(\mathcal O^\sigma_{mn}\bigr)^{-1}U^{\sigma\dagger}_m$ gives the expectation values of any electron operator by Wick's theorem. Its elements are \begin{equation}
\bigl(W^\sigma_{mn}\bigr)_{\ell\ell'}=\frac{\langle m|a^\dagger_{\ell'\sigma}a_{\ell\sigma}|n\rangle}{\langle m|n\rangle},
\end{equation}
where $\ell$ and $\ell'$ are site indices; the first index refers to the annihilation operator acting on $|n\rangle$ and the second to the creation operator, so that the order of the indices is opposite to that of the operators. In Eq.~\eqref{eq:HW} below, $W^\sigma_{\ell\ell'}$ stands for $(W^\sigma_{mn})_{\ell\ell'}$. For the coherent states, $\langle z_m|z_n\rangle=\exp\sum_k\bigl(-\tfrac12|z_{m,k}|^2-\tfrac12|z_{n,k}|^2+z^*_{m,k}z_{n,k}\bigr)$. The Hamiltonian matrix element is
\begin{align}
\frac{\langle m,z_m|H|n,z_n\rangle}{\langle m|n\rangle\langle z_m|z_n\rangle}\equiv H(W_{mn})
&=-\sum_{\ell\sigma}\bigl[1+\alpha X_{mn}(\ell)\bigr]\bigl(W^\sigma_{\ell+1,\ell}+W^\sigma_{\ell,\ell+1}\bigr)
+U\sum_\ell W^\uparrow_{\ell\ell}W^\downarrow_{\ell\ell}\nonumber\\
&\quad+V\sum_\ell\bigl(W^\uparrow_{\ell\ell}+W^\downarrow_{\ell\ell}\bigr)\bigl(W^\uparrow_{\ell+1,\ell+1}+W^\downarrow_{\ell+1,\ell+1}\bigr)\nonumber\\
&\quad-V\sum_{\ell\sigma}W^\sigma_{\ell+1,\ell}W^\sigma_{\ell,\ell+1}+\sum_k\omega(k)\Bigl(z^*_{m,k}z_{n,k}+\frac12\Bigr),
\label{eq:HW}
\end{align}
where
\begin{equation}
X_{mn}(\ell)=\sum_k\sqrt{\frac{1}{2MN\omega(k)}}\,\bigl(z_{n,k}+z^*_{m,-k}\bigr)e^{ik\ell}\bigl(1-e^{ik}\bigr)
\end{equation}
is the matrix element of $q_\ell-q_{\ell+1}$. Thus the electron and lattice degrees of freedom are coupled in each configuration through the hopping term, and all expectation values on a finite cluster, including those of the energy, the correlation functions, and the lattice distortion, are evaluated from the wave function.

The initial state is the static solution at the initial value of $V$. It is obtained by minimizing the energy $\langle\Phi|H|\Phi\rangle/\langle\Phi|\Phi\rangle$ with respect to the same parameters, $C_m$, $U^\sigma_m$, and $z_{m,k}$.

\subsection{Variational principle and equations of motion}

The time-dependent Schr\"odinger equation, $(i\hbar\,\partial_t-H)|\Phi\rangle=0$, is solved approximately by the McLachlan variational principle \cite{McLachlan,Martinazzo}. Let $\{x_i\}$ denote all the variational parameters. The time derivative of the wave function is $|\dot\Phi\rangle=\sum_i(\partial|\Phi\rangle/\partial x_i)\,\dot x_i$, and the error of the equation of motion is $|\epsilon\rangle=i\hbar|\dot\Phi\rangle-H|\Phi\rangle$. At time $t$ the parameters $x_i$, and hence $|\Phi\rangle$, are known; the unknowns are their rates of change $\dot x_i$, which are determined so that the norm $\langle\epsilon|\epsilon\rangle$ is minimal. Varying $\langle\epsilon|\epsilon\rangle$ with respect to $\dot x_i$ gives the McLachlan variational principle in the form
\begin{equation}
\mathrm{Im}\,\langle\delta\dot\Phi|\,i\hbar\dot\Phi-H\Phi\rangle=0,\qquad
|\delta\dot\Phi\rangle=\sum_i\frac{\partial|\Phi\rangle}{\partial x_i}\,\delta\dot x_i .
\label{eq:McL}
\end{equation}
For the coefficients $C_m$ and for the orbitals, the variation $\delta\dot x_i$ can be multiplied by the imaginary unit $i$, that is, $|\delta\dot\Phi\rangle\to i|\delta\dot\Phi\rangle$ is allowed. Then both the real and the imaginary parts of $\langle\delta\dot\Phi|\epsilon\rangle$ vanish, and Eq.~\eqref{eq:McL} reduces to
\begin{equation}
\langle\delta\dot\Phi|\,i\hbar\dot\Phi-H\Phi\rangle=0 .
\label{eq:McL2}
\end{equation}
Since $|\delta\dot\Phi\rangle$ is spanned by $\partial|\Phi\rangle/\partial x_i$, this states that $|\epsilon\rangle=i\hbar|\dot\Phi\rangle-H|\Phi\rangle$ is orthogonal to the tangent space spanned by $\partial|\Phi\rangle/\partial x_i$. The variation $|\delta\Phi\rangle=\sum_i(\partial|\Phi\rangle/\partial x_i)\,\delta x_i$ also lies in this space, so that
\begin{equation}
\langle\delta\Phi|\,i\hbar\dot\Phi-H\Phi\rangle=0 ,
\label{eq:DF}
\end{equation}
which is the Dirac--Frenkel condition \cite{Dirac}.

The coherent state, however, depends on $z^*$ as well as on $z$, through the normalization factor. Its derivatives are
\begin{equation}
\frac{\partial}{\partial z_k}|z\rangle=\Bigl(b^\dagger_k-\frac12z^*_k\Bigr)|z\rangle,\qquad
\frac{\partial}{\partial z^*_k}|z\rangle=-\frac12 z_k|z\rangle .
\end{equation}
The variation $\delta\dot z_k\to i\,\delta\dot z_k$ is accompanied by $\delta\dot z^*_k\to-i\,\delta\dot z^*_k$, and $|\delta\dot\Phi\rangle\to i|\delta\dot\Phi\rangle$ does not hold. We therefore cannot use Eq.~\eqref{eq:DF} for the coherent-state amplitudes $z_{m,k}$ and have to start from Eq.~\eqref{eq:McL} itself. The procedure is as follows.

(i) The wave function depends on both $z_{m,k}$ and $z^*_{m,k}$. Let $c$ denote the parameters on which $|\Phi\rangle$ depends only through $c$ and not through $c^*$, namely the coefficients $C_m$ and the orbital parameters $\Lambda^m_{\mu\alpha}$ (introduced below), and let $|\Phi_m\rangle=C_m|m,z_m\rangle$ be the $m$-th term of $|\Phi\rangle$. The variation is
\begin{equation}
|\delta\dot\Phi\rangle=\sum_c\frac{\partial|\Phi\rangle}{\partial c}\,\delta\dot c
+\sum_{m,k}\Bigl(\frac{\partial|\Phi_m\rangle}{\partial z_{m,k}}\,\delta\dot z_{m,k}+\frac{\partial|\Phi_m\rangle}{\partial z^*_{m,k}}\,\delta\dot z^*_{m,k}\Bigr),
\end{equation}
where, from the derivatives of the coherent state given above,
\begin{equation}
\frac{\partial|\Phi_m\rangle}{\partial z_{m,k}}=C_m\Bigl(b^\dagger_k-\frac12z^*_{m,k}\Bigr)|m,z_m\rangle,\qquad
\frac{\partial|\Phi_m\rangle}{\partial z^*_{m,k}}=-\frac12\,C_m\,z_{m,k}|m,z_m\rangle .
\end{equation}

(ii) We define
\begin{equation}
a_c=\Bigl\langle\frac{\partial\Phi}{\partial c}\Bigm|\epsilon\Bigr\rangle,\qquad
A_{m,k}=\Bigl\langle\frac{\partial\Phi_m}{\partial z_{m,k}}\Bigm|\epsilon\Bigr\rangle,\qquad
B_{m,k}=\Bigl\langle\frac{\partial\Phi_m}{\partial z^*_{m,k}}\Bigm|\epsilon\Bigr\rangle ,
\end{equation}
where $\langle\partial\Phi/\partial x|$ is the bra of the ket $\partial|\Phi\rangle/\partial x$ and $|\epsilon\rangle$ is the error of the whole wave function, so that the matrix elements are summed over the configurations $n$ of the ket. Then
\begin{equation}
\langle\delta\dot\Phi|\epsilon\rangle=\sum_c a_c\,\delta\dot c^{\,*}+\sum_{m,k}\bigl(A_{m,k}\,\delta\dot z^*_{m,k}+B_{m,k}\,\delta\dot z_{m,k}\bigr).
\end{equation}

(iii) The imaginary part is $2i\,\mathrm{Im}\,X=X-X^*$ with $X=\langle\delta\dot\Phi|\epsilon\rangle$. With $D_{m,k}\equiv A_{m,k}-B^*_{m,k}$ we obtain
\begin{equation}
\mathrm{Im}\,\langle\delta\dot\Phi|\epsilon\rangle=\sum_c\mathrm{Im}\bigl(a_c\,\delta\dot c^{\,*}\bigr)+\sum_{m,k}\mathrm{Im}\bigl(D_{m,k}\,\delta\dot z^*_{m,k}\bigr).
\end{equation}
This must vanish for arbitrary complex $\delta\dot c$ and $\delta\dot z_{m,k}$. A term $\mathrm{Im}(a_c\,\delta\dot c^*)$ vanishes for every complex $\delta\dot c$ only if $a_c=0$, as seen by considering $\delta\dot c$ and $i\,\delta\dot c$, which give $\mathrm{Im}\,(a_c\,\delta\dot c^*)=0$ and $\mathrm{Re}\,(a_c\,\delta\dot c^*)=0$, respectively. This is the Dirac--Frenkel condition. For the lattice, the same argument applied to $\mathrm{Im}(D_{m,k}\,\delta\dot z^*_{m,k})$ gives $D_{m,k}=0$, but $D_{m,k}=A_{m,k}-B^*_{m,k}$ contains $B_{m,k}$, which does not vanish because $|\Phi\rangle$ depends on $z^*_{m,k}$. The condition is therefore not of the Dirac--Frenkel form $A_{m,k}=0$. Omitting the index $m$ below because the equations for different $m$ have the same form, it reads
\begin{equation}
A_k-B^*_k=0 .
\label{eq:AB}
\end{equation}

(iv) Both $A_k$ and $B_k$ contain $|\epsilon\rangle$, that is, the time derivatives of all the parameters. $A_k$ contains $\dot C_n$, $\dot\Lambda$, $\dot z$, and $\dot z^*$, whereas $B^*_k$ contains their complex conjugates (see below). As a result, the Dirac--Frenkel equations for the coefficients and the orbitals and Eq.~\eqref{eq:AB} for the lattice coherent states, together with their complex conjugates, form a closed set of linear equations for $\dot C$, $\dot C^*$, $\dot\Lambda$, $\dot\Lambda^*$, $\dot z$, and $\dot z^*$, which are treated as independent unknowns.

We parametrize the time dependence of the orbitals by the Thouless transformation. Around the orbitals $|m\rangle$ at time $t$, the orbitals at the later time $t+\Delta t$ are
\begin{equation}
|m(t+\Delta t)\rangle=\exp\Bigl[\sum_{\mu\alpha}\bigl(\Lambda^m_{\mu\alpha}\,\tilde m^\dagger_\mu m_\alpha-\Lambda^{m*}_{\mu\alpha}\,m^\dagger_\alpha\tilde m_\mu\bigr)\Bigr]|m(t)\rangle ,
\label{eq:Thouless}
\end{equation}
where $m_\alpha$ ($\alpha=1,\dots,N_\sigma$) and $\tilde m_\mu$ ($\mu=1,\dots,N-N_\sigma$) are the occupied and the unoccupied orbitals of $|m(t)\rangle$, and $\Lambda^m_{\mu\alpha}=\Lambda^m_{\mu\alpha}(t+\Delta t)$. Since $\Lambda^m_{\mu\alpha}(t)=0$, the time derivative at $t$ is $\partial_t|m\rangle=\sum_{\mu\alpha}\dot\Lambda^m_{\mu\alpha}\tilde m^\dagger_\mu m_\alpha|m\rangle$, and the unknowns for the orbitals are the $\dot\Lambda^m_{\mu\alpha}$, $N_\sigma(N-N_\sigma)$ per spin and configuration. After each time step the orbitals are replaced by those of Eq.~\eqref{eq:Thouless}, which become the new expansion point.

The variation with respect to $C^*_m$ gives
\begin{equation}
\sum_nS_{mn}\Bigl\{i\dot C_n+C_n\Bigl[i\sum_{\mu\alpha}\dot\Lambda^n_{\mu\alpha}\frac{\langle m|n^\mu_\alpha\rangle}{\langle m|n\rangle}+iZ_{mn}-H(W_{mn})\Bigr]\Bigr\}=0,
\label{eq:Ceq}
\end{equation}
where $S_{mn}=\langle z_m|z_n\rangle\langle m|n\rangle$, $|n^\mu_\alpha\rangle=\tilde n^\dagger_\mu n_\alpha|n\rangle$ (for simplicity of notation the spin index of the determinants is omitted), and
\begin{equation}
Z_{mn}=\sum_k\Bigl(-\frac12z^*_{n,k}\dot z_{n,k}-\frac12z_{n,k}\dot z^*_{n,k}+z^*_{m,k}\dot z_{n,k}\Bigr)
\label{eq:Z}
\end{equation}
contains the time derivatives $\dot z_{n,k}$ and $\dot z^*_{n,k}$ of the coherent state.

For the orbitals, the variation of the bra orbitals at the later time gives the occupied--unoccupied components $\tilde u_m^\dagger(\cdots)u_m$, where $u_m$ and $\tilde u_m$ are the matrices of the occupied and the unoccupied orbitals of $\langle m|$ ($N\times N_\sigma$ and $N\times(N-N_\sigma)$), so that the number of equations equals the number of the unknowns $\dot\Lambda^m_{\mu\alpha}$ (the spin index is omitted):
\begin{align}
&\tilde u_m^\dagger\Bigl\{\sum_n\langle z_m|z_n\rangle\Bigl[\bigl\{(1-W_{mn})F(W_{mn})+H(W_{mn})-iZ_{mn}\bigr\}W_{mn}\langle m|n\rangle C_n-iW_{mn}\langle m|n\rangle\dot C_n\nonumber\\
&\qquad\qquad-i\sum_{\mu'\alpha'}\dot\Lambda^n_{\mu'\alpha'}\,W_{m,n^{\mu'}_{\alpha'}}\langle m|n^{\mu'}_{\alpha'}\rangle C_n\Bigr]\Bigr\}u_m=0 .
\label{eq:Leq}
\end{align}
Here $W_{m,n^{\mu'}_{\alpha'}}$ is the interstate density matrix between $\langle m|$ and $|n^{\mu'}_{\alpha'}\rangle$, $H(W)$ is given by Eq.~\eqref{eq:HW}, and $F(W)$ is the interstate Fock matrix, $F_{\ell'\ell}=\partial H(W)/\partial W_{\ell\ell'}$.

For the coherent-state amplitudes, $A_{m,k}$ and $B_{m,k}$ of Eq.~\eqref{eq:AB} are
\begin{align}
A_{m,k}&=C^*_m\sum_nS_{mn}\Bigl\{\Bigl(z_{n,k}-\frac12z_{m,k}\Bigr)\Bigl[i\dot C_n+C_n\Bigl(i\sum_{\mu\alpha}\dot\Lambda^n_{\mu\alpha}\frac{\langle m|n^\mu_\alpha\rangle}{\langle m|n\rangle}+iZ_{mn}-H(W_{mn})\Bigr)\Bigr]\nonumber\\
&\qquad\qquad+C_n\Bigl(i\dot z_{n,k}-\frac{\partial H(W_{mn})}{\partial z^*_{m,k}}\Bigr)\Bigr\},\nonumber\\
B_{m,k}&=-\frac12\,z^*_{m,k}\,C^*_m\,R_m ,
\label{eq:Aexp}
\end{align}
where $R_m$ is the left-hand side of Eq.~\eqref{eq:Ceq} and
\begin{equation}
\frac{\partial H(W_{mn})}{\partial z^*_{m,k}}=\omega(k)\,z_{n,k}-\alpha\sqrt{\frac{1}{2MN\omega(k)}}\,(1-e^{-ik})\sum_{\ell\sigma}e^{-ik\ell}\bigl(W^\sigma_{\ell+1,\ell}+W^\sigma_{\ell,\ell+1}\bigr).
\end{equation}
Equation~\eqref{eq:AB}, $A_{m,k}-B^*_{m,k}=0$, is the equation for the coherent-state amplitudes. Collecting the equations for the coefficients, the orbitals, and the coherent-state amplitudes, together with their complex conjugates required by Eq.~\eqref{eq:AB}, we obtain a linear system for the time derivatives,
\begin{equation}
\mathcal A\begin{pmatrix}\dot C\\ \dot C^*\\ \dot\Lambda\\ \dot\Lambda^*\\ \dot z\\ \dot z^*\end{pmatrix}=\begin{pmatrix}G_1\\ G_1^*\\ G_2\\ G_2^*\\ G_3\\ G_3^*\end{pmatrix}.
\label{eq:lin}
\end{equation}
Equation~\eqref{eq:lin} consists of six block equations. The first, third, and fifth blocks are the equations for the coefficients, Eq.~\eqref{eq:Ceq}, for the orbitals, Eq.~\eqref{eq:Leq}, and for the coherent-state amplitudes, Eq.~\eqref{eq:AB}; the second, fourth, and sixth blocks are their complex conjugates. The unknowns appear in the same pairs, $(\dot C,\dot C^*)$, $(\dot\Lambda,\dot\Lambda^*)$, and $(\dot z,\dot z^*)$. The two members of a pair are thus not independent. However, Eq.~\eqref{eq:AB} for the coherent-state amplitudes contains all of $\dot C$, $\dot C^*$, $\dot\Lambda$, $\dot\Lambda^*$, $\dot z$, and $\dot z^*$, so that $\dot z$ and $\dot z^*$ cannot be obtained separately from the other unknowns, and the correct solution is obtained only if the whole system is solved at once. Solving Eq.~\eqref{eq:lin} at a given time gives the time derivatives of $C$, $\Lambda$, and $z$; the derivatives of the complex conjugates are not integrated, since they are the complex conjugates of those of $(C,\Lambda,z)$.

\subsection{Time integration}

Equation~\eqref{eq:lin} is solved at each stage of the standard fourth-order Runge--Kutta method. At the beginning of each step, $\bm y=(C,\Lambda,z)$ is given by $C_m$ and $z_{m,k}$ at the time $t_n$ and $\Lambda=0$; at $t=0$ the configurations are those of the initial state, the static solution at the initial value of $V$. The kets in the matrix elements are always the configurations at the beginning of the step, that is, the coefficients, the orbitals, and the coherent states at $t_n$. The bras are the configurations at the stage: in the first stage they coincide with the kets, and in the second, third, and fourth stages they are constructed from the intermediate values $\bm y_n+\frac12\Delta t\,\bm k_1$, $\bm y_n+\frac12\Delta t\,\bm k_2$, and $\bm y_n+\Delta t\,\bm k_3$, where $\bm k_i$ is the time derivative of $\bm y$ obtained by solving Eq.~\eqref{eq:lin} in the $i$-th stage and $\Delta t$ is the time step. At the end of the step, $\bm y_{n+1}=\bm y_n+\frac{\Delta t}{6}(\bm k_1+2\bm k_2+2\bm k_3+\bm k_4)$. The time dependence of $V(t)$ is evaluated at the time of each stage. The time step is chosen so small that the results do not depend on it (see, for example, Sec.~\ref{sec:results} for $N=6$).

\subsection{Reconstruction of the configuration set}\label{sec:recon}

Because the configurations are nonorthogonal, the matrix $\mathcal A$ of Eq.~\eqref{eq:lin} contains the overlaps between them. When two configurations become close to each other, the overlap matrix becomes nearly singular, and Eq.~\eqref{eq:lin} becomes ill-conditioned: a small error in the right-hand side of Eq.~\eqref{eq:lin} produces a large change of the time derivatives, and the integration becomes unstable. This is the well-known difficulty of expansions in moving, nonorthogonal basis functions mentioned in Sec.~\ref{sec:intro}, for which several remedies have been proposed \cite{JoubertDoriol,Werther,Asaad,Hojlund,Schrader}.

We stabilize the calculation by a procedure that differs from these remedies and is used only when it is needed. At a time $t_{\rm r}$ at which the overlap matrix becomes ill-conditioned, we replace the configuration set by a new one, whose configurations are kept apart from each other: the largest normalized overlap between the configurations is limited by a cap, and the smallest eigenvalue of the normalized overlap matrix is kept above a threshold. The coefficients of the new configurations are then chosen so that the new state is as close as possible to the old state: the old state is projected onto the space spanned by the new configurations, under the constraint that the energy is unchanged. The new state is thus almost identical to the old state: in all cases used in this work, the fidelity between the two states exceeds 0.998, the energy is unchanged, and the correlation functions change by at most $4\times10^{-4}$ (Appendix~A). The equations of motion, Eq.~\eqref{eq:lin}, are not modified, and the time evolution is restarted from the new set; only the representation of the state at the time $t_{\rm r}$ is changed. In contrast to the reoptimization at every time step, the reconstruction is performed only once or twice in a run (Appendix~A lists all the reconstructions used in this work). The cost function, the optimization procedure, and the accuracy are described in Appendix~A, and the effect on the results is examined for $N=14$ in Sec.~\ref{sec:results}.

\section{Results}\label{sec:results}

\subsection{Setup and validation}\label{sec:valid}

In this section, we test TD-ResHF against exact time evolution and then apply it to $N=14$, a size beyond exact diagonalization. The interaction $V$ is raised linearly from an initial value $V_i$ to $V_f=2.5$ during a ramp of duration $\tau$ and is kept at $V_f$ afterwards,
\begin{equation}
V(t)=\begin{cases}V_i+(V_f-V_i)\,t/\tau & (0\le t\le\tau),\\ V_f & (t>\tau).\end{cases}
\end{equation}
We use $\tau=5$, $13$, and $30$ with $V_i=1.5$, and $\tau=100$ with $V_i=2.0$, so that the ramp rate $(V_f-V_i)/\tau$ is 0.2, 0.077, and 0.033 for $\tau=5$, 13, and 30, and 0.005 for $\tau=100$; the ramp with $\tau=100$ is the slowest in rate as well as in duration. The initial state is the static solution at $V_i$. We use five configurations (NBM$=5$) unless stated otherwise; a comparison with seven configurations is given in Sec.~\ref{sec:nbm}. The observables $N(\pi)$, $S(\pi)$, $B(\pi)$, and $D$ are defined in Sec.~\ref{sec:model}.

\begin{figure*}[!b]
\centering
\includegraphics[width=0.95\textwidth]{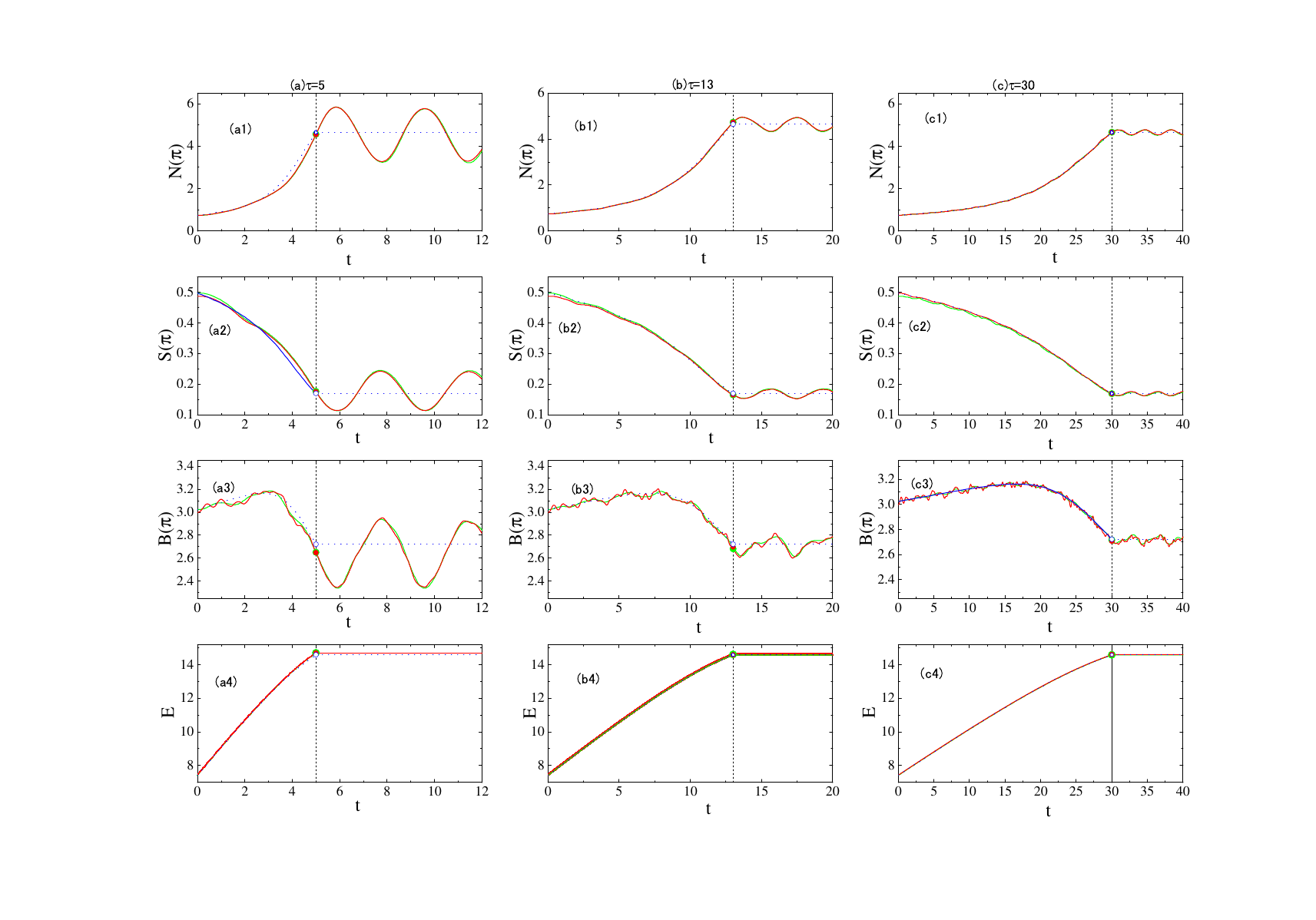}
\caption{(Color online) Electrons only, $N=10$: $N(\pi)$, $S(\pi)$, $B(\pi)$, and the energy $E$ for the ramps $\tau=5$ (a), $13$ (b), and $30$ (c). Red: TD-ResHF (NBM$=5$); green: exact dynamics; blue dashed: exact static values for the instantaneous $V(t)$; the dashed vertical line marks the end of the ramp. The circles mark the values at the end of the ramp: open, static; green, exact; red, TD-ResHF.}
\label{fig:el}
\end{figure*}

In what follows the deviation of TD-ResHF from the exact solution is given as the rms of the difference over the whole time range, divided by the time average of the exact value. For the electrons alone ($N=10$), TD-ResHF follows the exact time evolution for $\tau=5$, $13$, and $30$ (Fig.~\ref{fig:el}), including the oscillation after the ramp, with deviations within 1\%.

Table~\ref{tab:post} lists the excess energy $\Delta E=E-E_{\rm gs}(V=2.5)$ after the ramp, the amplitude $A_N$ of the oscillation of $N(\pi)$, and the averages of the correlation functions over the oscillation, for TD-ResHF and the exact solution. 
\begin{table}[!tb]
\caption{Excess energy $\Delta E=E-E_{\rm gs}(V=2.5)$ after the ramp, amplitude $A_N$ (half of the peak-to-peak value) of $N(\pi)$, and averages of $N(\pi)$, $S(\pi)$, and $B(\pi)$ over $\tau\le t\le\tau+7.5$ for $N=10$: TD-ResHF, with the exact values in parentheses. $A_N$ and the averages are in units of the static values at $V=2.5$, which are $N(\pi)=4.647$, $S(\pi)=0.1698$, $B(\pi)=2.722$ (electrons only) and $4.569$, $0.1729$, $2.849$ (with the lattice, $\pi$-mode model). $\Delta E$ is measured from the static energy at $V=2.5$ of the respective method (TD-ResHF: 14.5882 for the electrons only and 14.5578 with the lattice).}
\label{tab:post}
\begin{ruledtabular}
\begin{tabular}{llccccc}
 & $\tau$ & $\Delta E$ & $A_N/N_{\rm st}$ & $\langle N\rangle/N_{\rm st}$ & $\langle S\rangle/S_{\rm st}$ & $\langle B\rangle/B_{\rm st}$ \\
\hline
Electrons only & 5  & 0.114 (0.107)  & 0.275 (0.283) & 0.984 (0.981) & 1.046 (1.057) & 0.973 (0.974) \\
               & 13 & 0.016 (0.0061) & 0.066 (0.067) & 1.002 (0.999) & 0.995 (1.003) & 0.996 (0.998) \\
               & 30 & 0.011 (0.0010) & 0.026 (0.028) & 1.002 (1.000) & 0.992 (1.000) & 0.998 (1.000) \\
With lattice   & 5  & 0.189 (0.162)  & 0.195 (0.201) & 0.831 (0.830) & 1.161 (1.175) & 1.462 (1.448) \\
               & 13 & 0.135 (0.115)  & 0.067 (0.073) & 0.831 (0.834) & 1.160 (1.168) & 1.467 (1.450) \\
\end{tabular}
\end{ruledtabular}
\end{table}
The adiabaticity of the electrons, that is, whether they remain in the ground state for the instantaneous $V$ during the ramp, can be judged from these quantities. For the electrons alone, $\Delta E$ of TD-ResHF decreases from 0.114 at $\tau=5$ to 0.016 at $\tau=13$ and 0.011 at $\tau=30$ (exact: 0.107, 0.0061, and 0.0010), and the amplitude of the oscillation of $N(\pi)$, relative to the static value, from 28\% to 7\% and 3\%. For $\tau\ge13$ the excess energy of TD-ResHF is larger than the exact one by about 0.01, which is 0.1--0.2\% of the energy supplied by the ramp (about 7.2), and it does not affect the oscillation of the correlation functions, whose amplitude and period agree with the exact ones. The averages over the oscillation approach the static values: they differ from them by up to 5\% for $\tau=5$ and by 1\% or less for $\tau\ge13$. The electrons are therefore clearly nonadiabatic at $\tau=5$, nearly adiabatic at $\tau=13$, and adiabatic at $\tau=30$, where the correlation functions follow the static values for the instantaneous $V$. The origin of the oscillation is the following. After the ramp the electrons are mainly in the ground state at $V=2.5$ with a small admixture of the lowest excited state that couples to $N(\pi)$. The oscillation is the interference between the two states, and its frequency is the excitation energy. Because the interference term is proportional to the probability amplitude of the excited state rather than to its weight, a weight of a few tenths of a percent already gives a visible oscillation. The quantitative analysis is given in Appendix~\ref{app:amp}.

\begin{figure*}[t]
\centering
\includegraphics[width=0.85\textwidth]{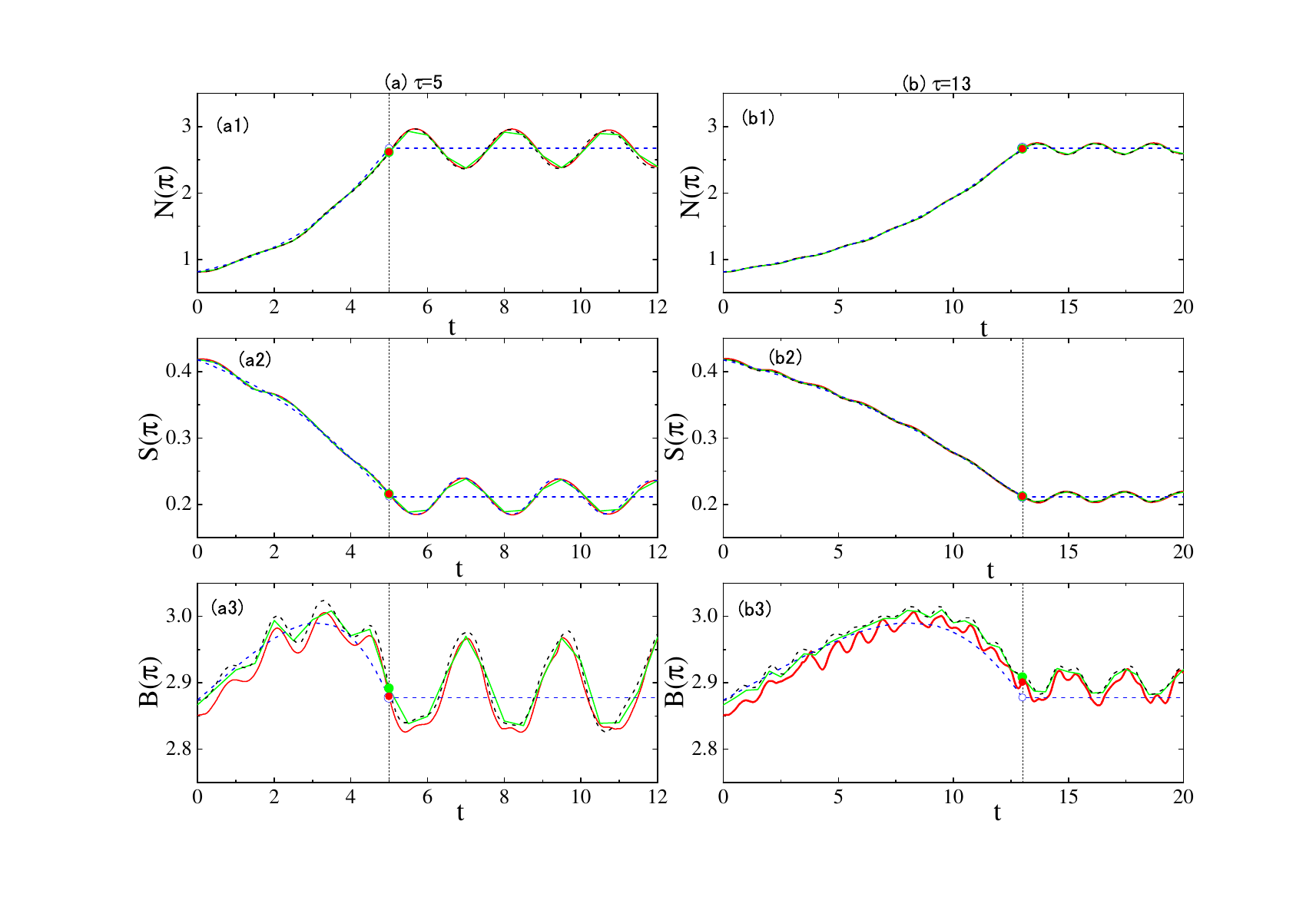}
\caption{(Color online) Electron--lattice system, $N=6$: $N(\pi)$, $S(\pi)$, and $B(\pi)$ for $\tau=5$ (a) and $13$ (b). Green: exact diagonalization with all lattice modes; black dashed: exact solution restricted to the $\pi$ mode; red: TD-ResHF (NBM$=5$); blue dashed: exact static value for the instantaneous $V(t)$. The circles mark the values at the end of the ramp (dashed vertical line): open, static; green, exact diagonalization with all modes; red, TD-ResHF.}
\label{fig:n6}
\end{figure*}

\begin{figure*}[t]
\centering
\includegraphics[width=0.95\textwidth]{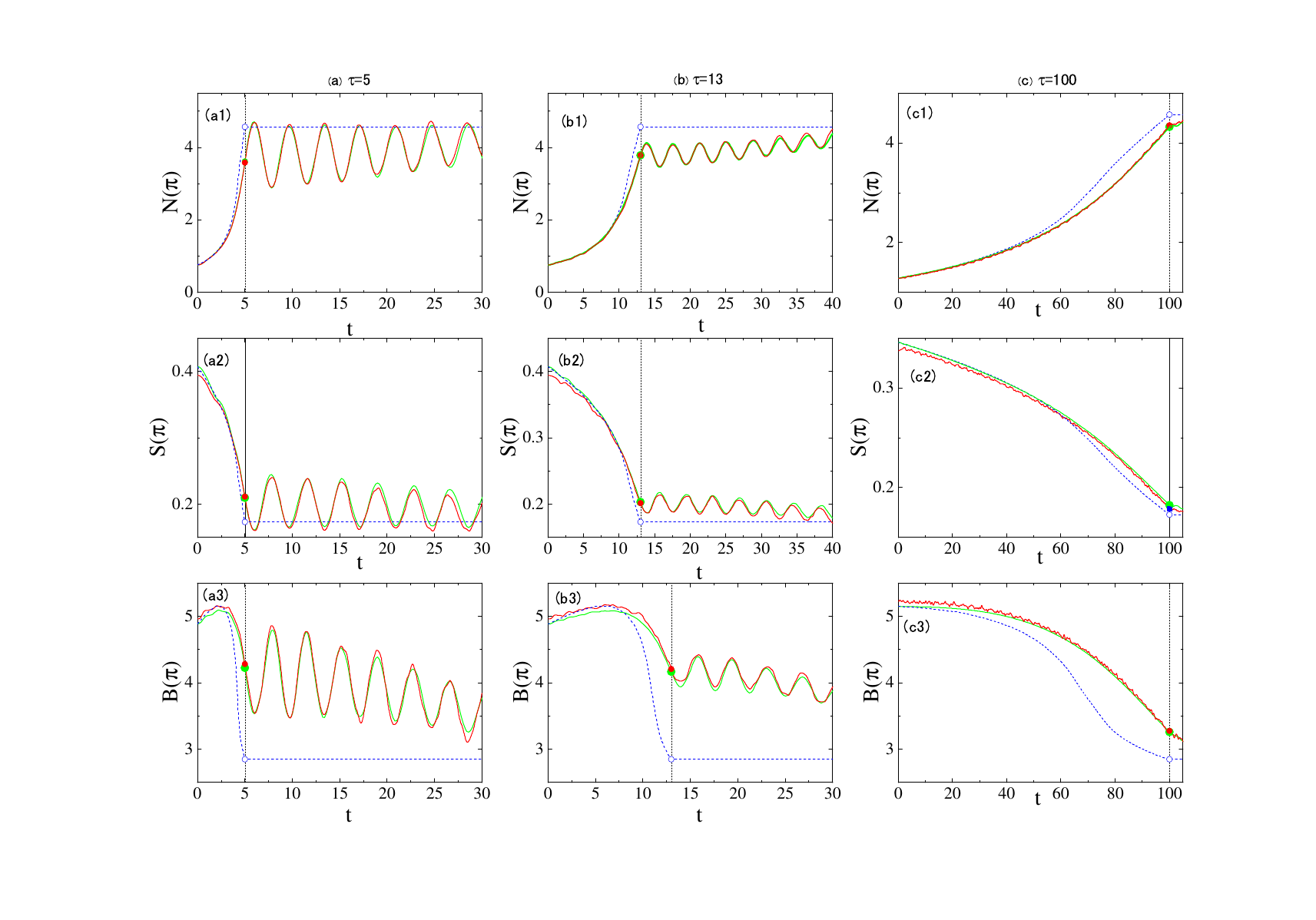}
\caption{(Color online) Electron--lattice system, $N=10$: $N(\pi)$, $S(\pi)$, and $B(\pi)$ for $\tau=5$ (a), $13$ (b), and $100$ (c). Green: exact solution restricted to the $\pi$ phonon mode; red: TD-ResHF; blue dashed: exact static value for the instantaneous $V(t)$. The ramp for $\tau=100$ is from $V=2.0$ to $2.5$. The circles mark the values at the end of the ramp (dashed vertical line): open, static; green, exact; red, TD-ResHF.}
\label{fig:corr}
\end{figure*}

\begin{figure*}[t]
\centering
\includegraphics[width=0.85\textwidth]{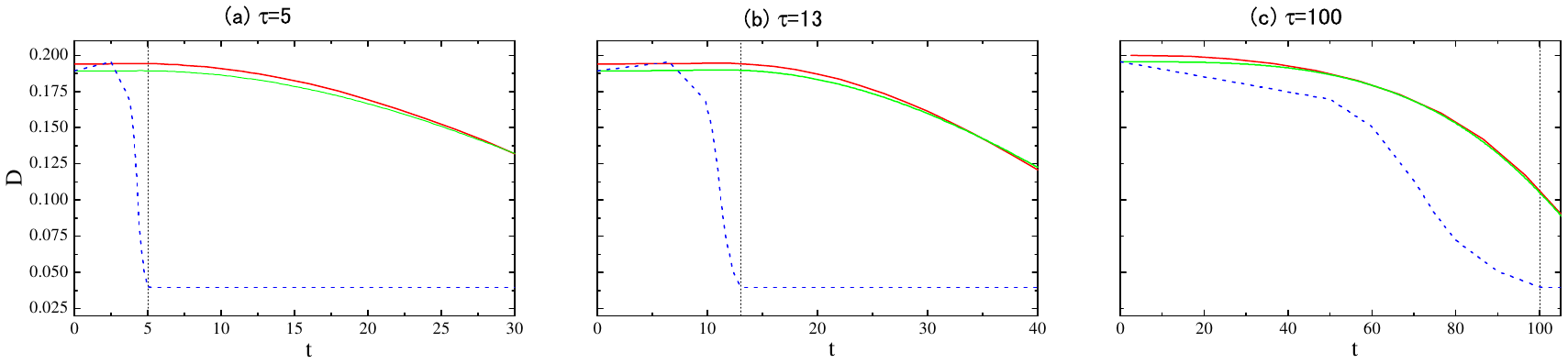}
\caption{(Color online) Dimerization amplitude $D$ for $N=10$ and $\tau=5$ (a), $13$ (b), and $100$ (c). Green: exact dynamics restricted to the $\pi$ mode; red squares: TD-ResHF; blue dashed: exact static value $D(V(t))$ for the instantaneous $V$. The dashed vertical line marks the end of the ramp.}
\label{fig:D}
\end{figure*}

We now compare with exact solutions of the electron--lattice system. For $N=6$ the exact solution is obtained by exact diagonalization of the full Hamiltonian, including all lattice modes (Appendix~\ref{app:ed}). The TD-ResHF results agree with it for $\tau=5$ and $13$ (Fig.~\ref{fig:n6}), with deviations below 1\%. $B(\pi)$ appears to deviate more in the figure, but the rms deviation is 0.7--0.8\%, and it is mostly a constant offset already present in the initial state, which comes from the static ResHF solution. For $N=10$ the exact diagonalization with all lattice modes is not feasible. As the reference we therefore use the exact time evolution of the model in which the lattice is restricted to the $\pi$ mode. For $N=6$ this restricted solution differs from the all-mode solution by 2--7\% at most (black dashed lines in Fig.~\ref{fig:n6}), so that it is a good reference. The TD-ResHF results follow it for $\tau=5$, $13$, and $100$ (Fig.~\ref{fig:corr}). The deviation is within 1--2\%, so that TD-ResHF gives reliable results also for $N=10$. The agreement includes the lattice degrees of freedom: the dimerization amplitude $D(t)$ also follows the exact solution (Fig.~\ref{fig:D}). With this agreement established, we analyze the dynamics after the ramp for $N=6$ and $10$ (Sec.~\ref{sec:static}) and for the slowest ramp (Sec.~\ref{sec:slow}), and then treat $N=14$, which is beyond the reach of exact diagonalization, with TD-ResHF (Sec.~\ref{sec:n14}).

\clearpage
\subsection{Approach to the static state and the frozen lattice}\label{sec:static}

We analyze here the TD-ResHF results for $N=6$ and $10$, which agree with the exact solution (Sec.~\ref{sec:valid}), to establish the picture of the dynamics. This picture is then extended to $N=14$, a size beyond exact diagonalization, where the configurations of the wave function are also examined (Sec.~\ref{sec:n14}). In the static state, $N(\pi)$ increases and $S(\pi)$ decreases as $V$ is raised, reflecting the change from spin-density-wave to charge-density-wave correlations, while $B(\pi)$ first increases and then decreases (the static curves are the dashed lines in Figs.~\ref{fig:el}, \ref{fig:n6}, and \ref{fig:corr}). In the dynamics, the growth of $N(\pi)$ and the decrease of $S(\pi)$ lag behind the static ones when the lattice does not follow the ramp. After the ramp the correlation functions oscillate. The question is how far the system is from the static state at the instantaneous $V(t)$ (dashed lines in Figs.~\ref{fig:el}, \ref{fig:n6}, and \ref{fig:corr}). The circles mark the values at the end of the ramp, and Table~\ref{tab:end} lists their deviations from the static values.

\paragraph*{Electrons only.} The state at the end of the ramp is close to the static one already for $\tau=5$, and the oscillation after the ramp is centered on the static value (Table~\ref{tab:post}). The electrons alone therefore follow the ramp; the size of the oscillation decreases as $\tau$ increases.

\paragraph*{Electron--lattice system, $N=6$.} The effect of the lattice is small, because the static dimerization $D$ is small even at $V=1.5$: it is 0.05 in the static ResHF solution and 0.06 in the exact solution. The Peierls-type transition is hardly visible at $N=6$ owing to the finite-size effect, and the static values of the correlation functions (dashed lines in Fig.~\ref{fig:n6}) change only smoothly as $V$ is raised from 1.5 to 2.5. After the ramp, $N(\pi)$ and $S(\pi)$ oscillate around the static values at $V=2.5$, as for the electrons alone (Fig.~\ref{fig:n6}), and at the end of the ramp the correlation functions deviate from the static values by only a few percent (Table~\ref{tab:end}). The oscillation of $B(\pi)$ is also centered close to the static value, but slightly above it, by about 1\%. The reason is the lattice. It keeps a small distortion from the initial state, larger than that of the static solution at $V=2.5$, and in a more dimerized lattice the bond order is larger, so that the electrons settle at a $B(\pi)$ above the static value. The distortion is small at $N=6$, so that the effect on $N(\pi)$ and $S(\pi)$ is hardly visible.

\paragraph*{Electron--lattice system, $N=10$.} Here the behavior is different. The Peierls-type transition of the static state is clear: between $V=2.25$ and $2.5$ the dimerization of the static solution drops and the static values of $N(\pi)$ and $B(\pi)$ change abruptly. The correlation functions of the dynamics do not follow this change (Fig.~\ref{fig:corr}). At the end of the ramp the correlation functions deviate from the static values by 17--48\% for $\tau=5$ and $13$ (Table~\ref{tab:end}), and the oscillation after the ramp is not centered on the static value: it is centered below it for $N(\pi)$ and above it for $S(\pi)$ and $B(\pi)$. The deviation is smaller for $\tau=100$ (Fig.~\ref{fig:corr}(c)), because the ramp is much slower. In contrast, the deviations are only a few percent for the electrons alone and for $N=6$ (Table~\ref{tab:end}), so that the large deviations at $N=10$ are caused by the lattice. The excess energy of the electrons points in the same direction: for $\tau=13$ it is much larger with the lattice than for the electrons alone (Table~\ref{tab:post}).

\begin{table}[t]
\caption{Deviation of $N(\pi)$, $S(\pi)$, and $B(\pi)$ at the end of the ramp from the static values at $V=2.5$, in percent: TD-ResHF (left) and exact solution (right) in each column. The ramp for $\tau=100$ is from $V=2.0$ to $2.5$.}
\label{tab:end}
\begin{ruledtabular}
\begin{tabular}{llrrr}
 & $\tau$ & $N(\pi)$ & $S(\pi)$ & $B(\pi)$ \\
 &  & TD / exact & TD / exact & TD / exact \\
\hline
Electrons only, $N=10$ & 5   & $-1.7\,/\,-2.1$ & $+3.1\,/\,+4.2$ & $-2.6\,/\,-2.7$ \\
                       & 13  & $+1.8\,/\,+1.8$ & $-3.0\,/\,-2.4$ & $-1.5\,/\,-0.9$ \\
                       & 30  & $+0.2\,/\,+0.3$ & $-0.2\,/\,-0.4$ & $-0.2\,/\,-0.6$ \\
With lattice, $N=6$    & 5   & $-2.1\,/\,-2.3$ & $+2.0\,/\,+2.1$ & $+0.1\,/\,+0.5$ \\
                       & 13  & $-0.9\,/\,-0.5$ & $+0.5\,/\,+0.4$ & $+0.2\,/\,+1.1$ \\
With lattice, $N=10$   & 5   & $-22\,/\,-21$   & $+22\,/\,+21$   & $+50\,/\,+48$ \\
                       & 13  & $-17\,/\,-17$   & $+17\,/\,+17$   & $+48\,/\,+46$ \\
                       & 100 & $-5\,/\,-5$     & $+4\,/\,+6$     & $+15\,/\,+14$ \\
\end{tabular}
\end{ruledtabular}
\end{table}

For $\tau=5$ and $13$ the lattice does not follow the ramp, as seen in the dimerization amplitude $D(t)$ (Fig.~\ref{fig:D}(a),(b)). If the lattice followed the ramp, that is, relaxed to the static state at each $V$, $D$ would drop to the small value of the static solution at the end of the ramp (dashed lines). In the dynamics, $D$ stays at about 0.19 until the end of the ramp, against the static value 0.040 at $V=2.5$, and decreases only slowly afterwards. The ramp time is much shorter than the phonon period $2\pi/\omega\approx95$, so the lattice remains frozen near its initial distortion while the electrons move, and the charge and spin structure develops under this distortion.

The results for $N=10$ can be summarized as follows. For the electrons alone, the correlation functions approach the static values as $\tau$ increases: the electrons are clearly nonadiabatic for $\tau=5$ and nearly adiabatic for $\tau=13$ and longer (Table~\ref{tab:post}). With the lattice, both subsystems matter. For $\tau=5$ neither the electrons nor the lattice follow the ramp, and the deviation from the static values is expected. For $\tau=13$ the electrons are nearly adiabatic, but the ramp is much shorter than the phonon period, the lattice stays frozen at its initial distortion, and the correlation functions stay away from the static values (Table~\ref{tab:end}). The difference between $\tau=5$ and $13$ is therefore the size of the oscillation after the ramp, which is determined by the electrons (the oscillation has the frequency of the electronic excitation, much higher than the phonon frequency). The centers of the oscillation are nearly the same for $\tau=5$ and $13$ and are determined by the frozen lattice. For $\tau=100$, comparable to the phonon period, the freezing is largely released and $D$ decreases during the ramp (Fig.~\ref{fig:D}(c)), but it does not reach the static value; the correlation functions are much closer to the static ones than for the other ramp times (Fig.~\ref{fig:corr}(c)), but not equal to them. In all three cases the lattice has not relaxed to the static state, and the electrons therefore do not reach the static state of the coupled system. The deviation from the static state thus has two parts: the oscillation after the ramp, which decreases as the electrons become adiabatic, and the offset of the center of the oscillation from the static value, which is caused by the lattice distortion that remains from the initial state and decreases only as the lattice relaxes. The latter is still present for the slowest ramp, $\tau=100$, which is described in Sec.~\ref{sec:slow}.

\clearpage
\subsection{The slowest ramp: the lattice stays behind}\label{sec:slow}

For $\tau=100$, comparable to the phonon period, one might expect the lattice to follow the ramp. The electrons follow such a slow ramp adiabatically: without the lattice they already do so for $\tau\le30$ (Table~\ref{tab:post}), and in the coupled system the oscillation of $N(\pi)$ with the electronic period ($\approx3.7$) is below 1\% of its value in the short time after the ramp (Fig.~\ref{fig:corr}(c)). $D$ remains well above its static value (Fig.~\ref{fig:D}(c)). The lattice stays dimerized for most of the ramp, and $D$ decreases appreciably only near its end. In $N(\pi)$ this appears as a delay with respect to the static value: the charge correlation lags behind the static value, and the delay grows to about 11\% at $V=2.4$ and shrinks to 5\% at the end of the ramp, $V=2.5$. The delay is not due to excitations of the electrons themselves, since the electrons follow this ramp adiabatically. It is a consequence of the lattice distortion that does not follow the ramp. The electrons move under a distortion that competes with the charge-density wave: the charge correlation still grows, but less than in the static state, while the bond-order correlation $B(\pi)$ is enhanced.

\subsection{Dependence on the number of configurations}\label{sec:nbm}

All results above use five configurations (NBM$=5$). To see whether more configurations change the results, we repeated the calculation for $N=10$ and $\tau=5$ with seven configurations (NBM$=7$), starting from the static solution with seven configurations at $V=1.5$ and following the evolution to $t=12$. Since the seven configurations are already close to each other in the initial state, the set was reconstructed (Appendix~\ref{app:recon}) at the start of the evolution ($t=0.25$) and again at $t=5.5$.

\begin{figure}[t]
\centering
\includegraphics[width=3.4in]{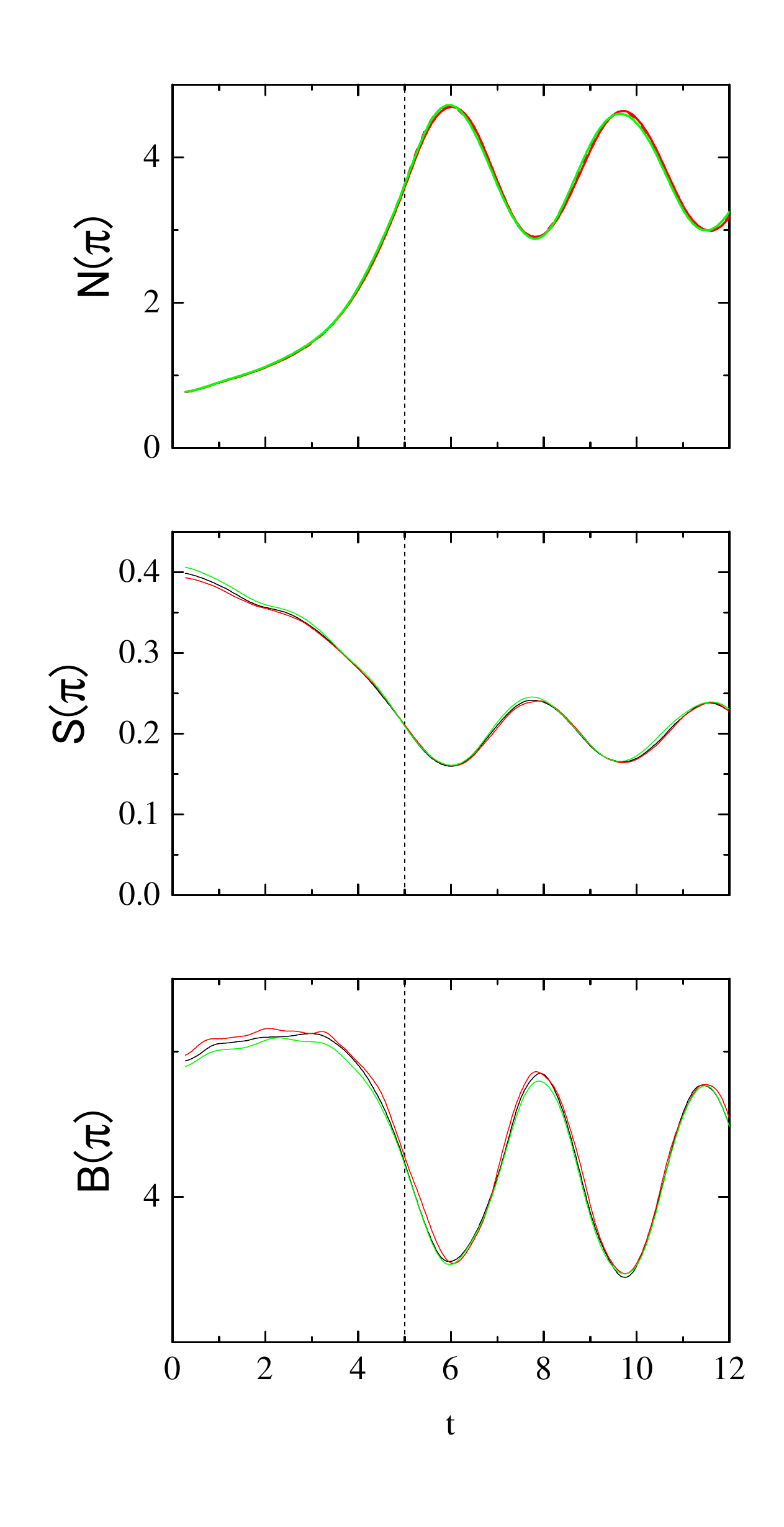}
\caption{(Color online) $N(\pi)$, $S(\pi)$, and $B(\pi)$ for $N=10$ and $\tau=5$ with five (red) and seven (black) configurations, compared with the exact solution restricted to the $\pi$ mode (green). The dashed vertical line marks the end of the ramp.}
\label{fig:nbm}
\end{figure}

Figure~\ref{fig:nbm} compares NBM$=7$ and NBM$=5$ with the exact solution restricted to the $\pi$ mode. The two TD-ResHF results are close to each other for all three quantities, with an rms difference of about 1\% of the mean value of the exact solution, and the energies differ by 0.1--0.2\%. The results with NBM$=7$ are slightly closer to the exact solution: the rms deviation is 0.7--1.3\% against 1.4--2.0\% with NBM$=5$. The five configurations used here are therefore sufficient for the quantities discussed.

\subsection{$N=14$: a size beyond exact diagonalization}\label{sec:n14}

\begin{figure}[!t]
\centering
\includegraphics[width=3.4in]{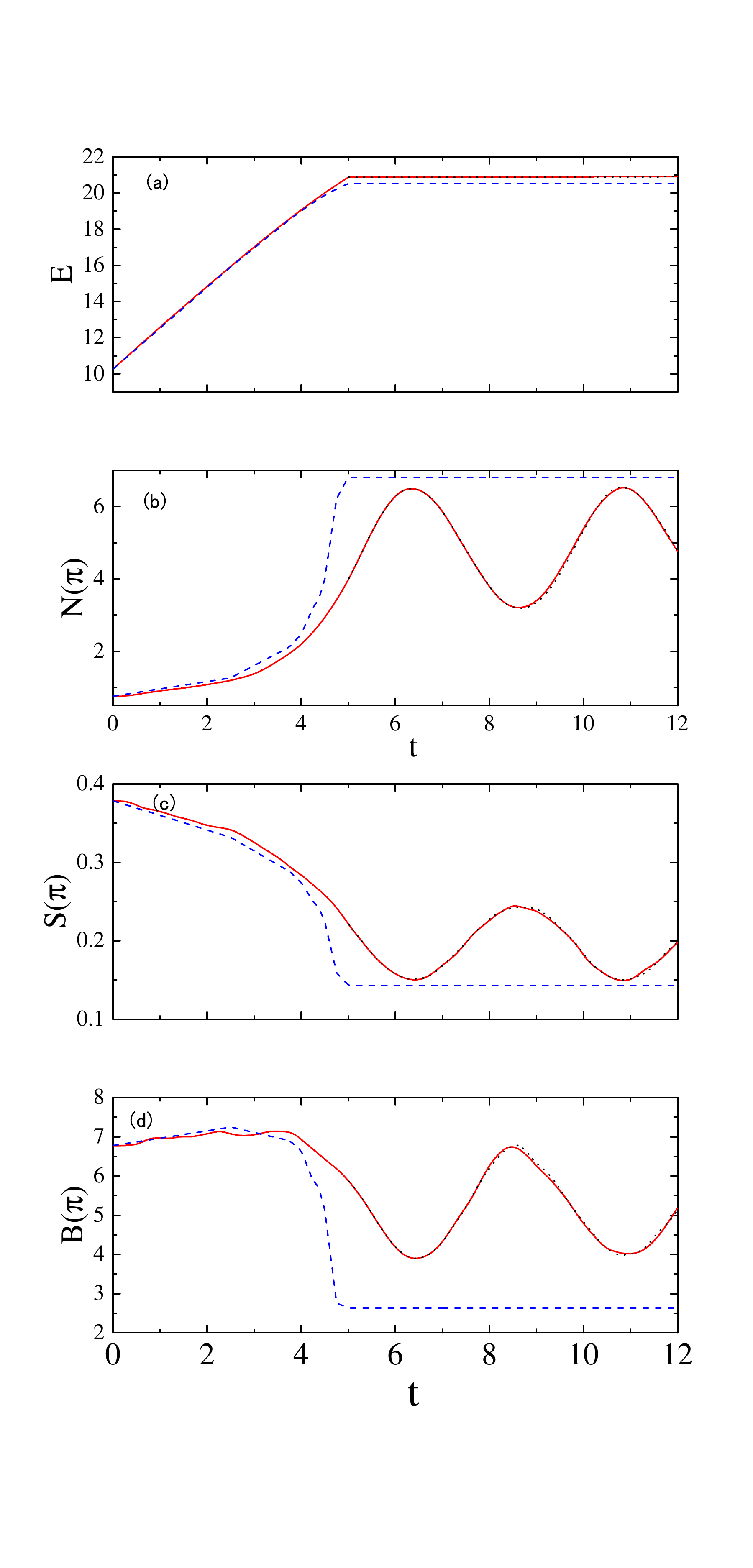}
\caption{(Color online) $N=14$, $\tau=5$: (a) energy $E$, (b) $N(\pi)$, (c) $S(\pi)$, and (d) $B(\pi)$. Red solid: TD-ResHF (restart after $t=5$); black dotted: without reconstruction; blue dashed: static ResHF solution at the instantaneous $V(t)$. The dashed vertical line marks the end of the ramp.}
\label{fig:corr14}
\end{figure}

\begin{figure}[!t]
\centering
\includegraphics[width=3.4in]{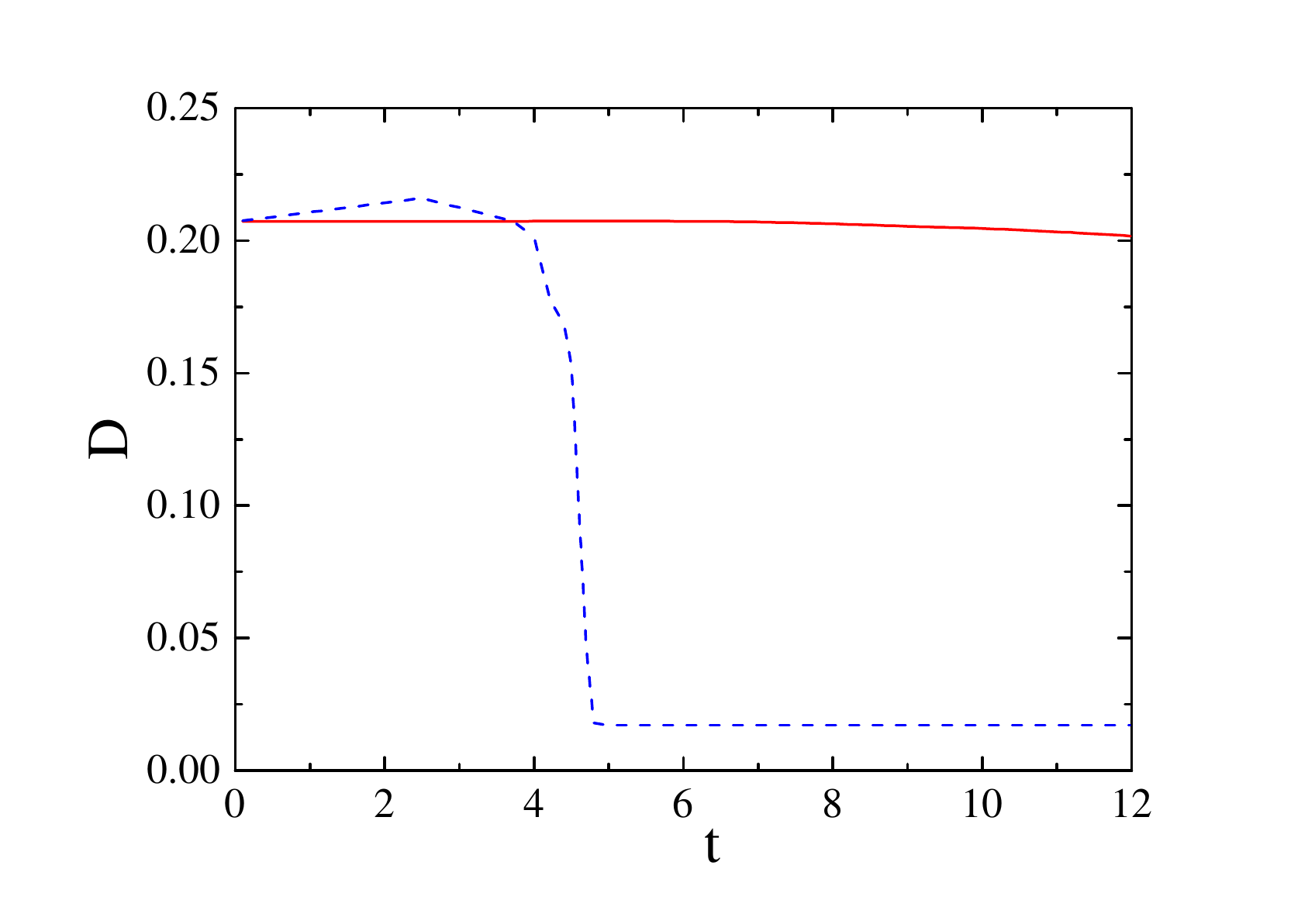}
\caption{(Color online) Dimerization amplitude $D$ for $N=14$, $\tau=5$. Red solid: TD-ResHF dynamics; blue dashed: static ResHF solution at the instantaneous $V(t)$.}
\label{fig:D14}
\end{figure}

\begin{figure*}[!t]
\centering
\includegraphics[width=0.48\textwidth]{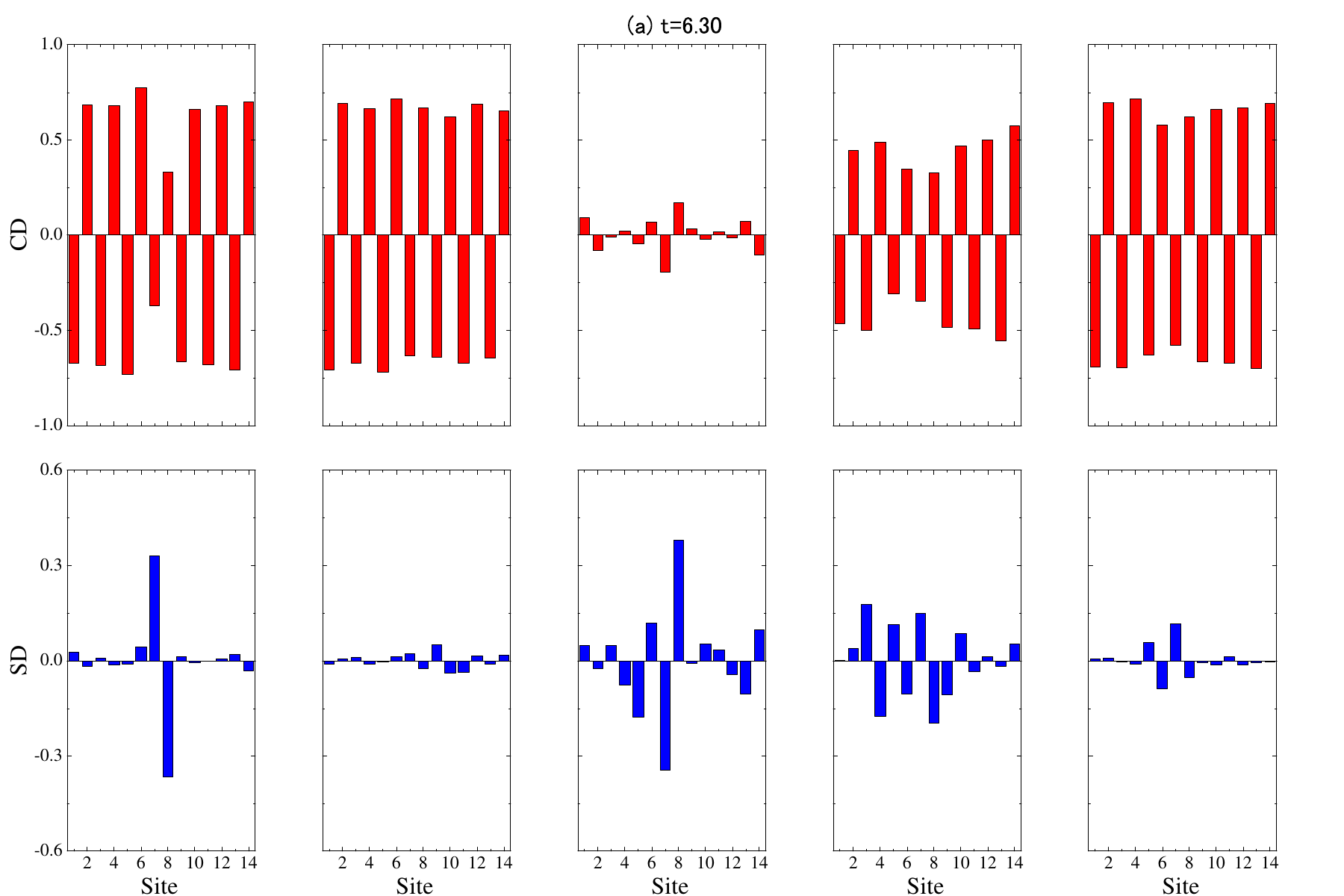}\hfill
\includegraphics[width=0.48\textwidth]{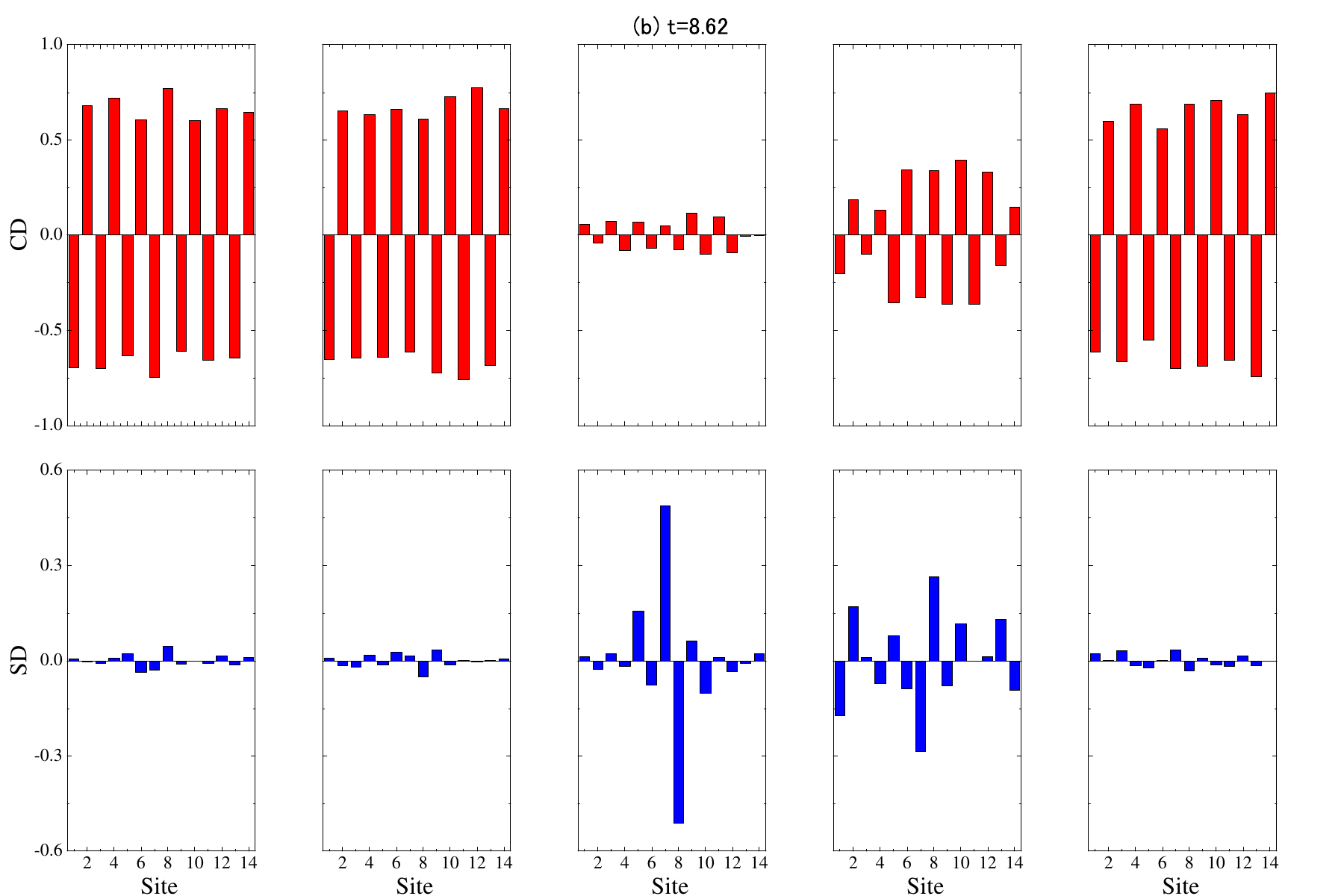}\\[4pt]
\includegraphics[width=0.48\textwidth]{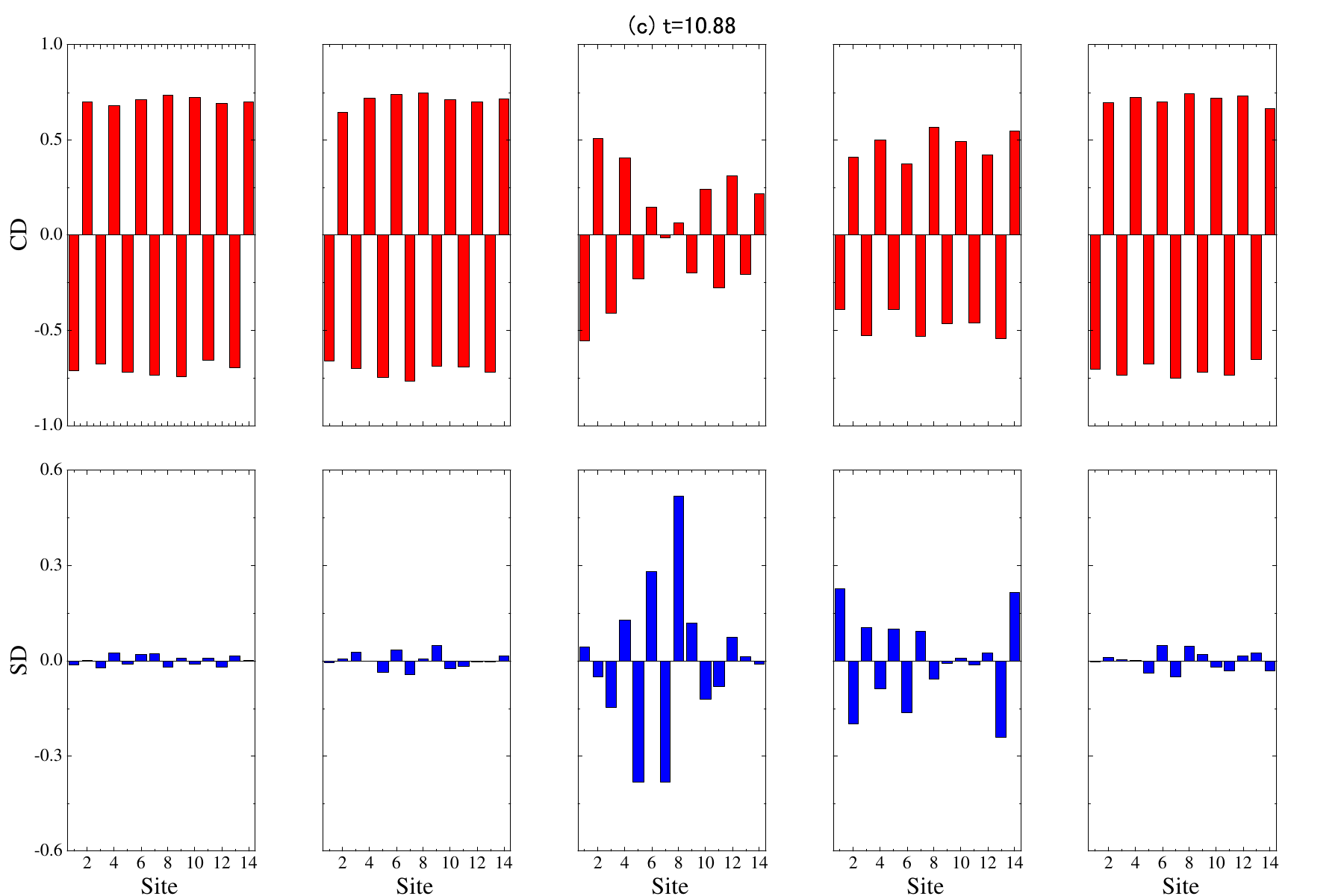}
\caption{(Color online) Configurations of the $N=14$ wave function at the maximum ($t=6.30$, a), at the minimum ($t=8.62$, b), and at the next maximum ($t=10.88$, c) of $N(\pi)$. For each of the five configurations, the charge $n_\uparrow+n_\downarrow$ (CD, shown relative to 1) and the spin $n_\uparrow-n_\downarrow$ (SD, relative to 0) at each site. The shapes of the configurations are almost the same at the maxima and at the minimum.}
\label{fig:conf}
\end{figure*}

\begin{figure}[!t]
\centering
\includegraphics[width=3.4in]{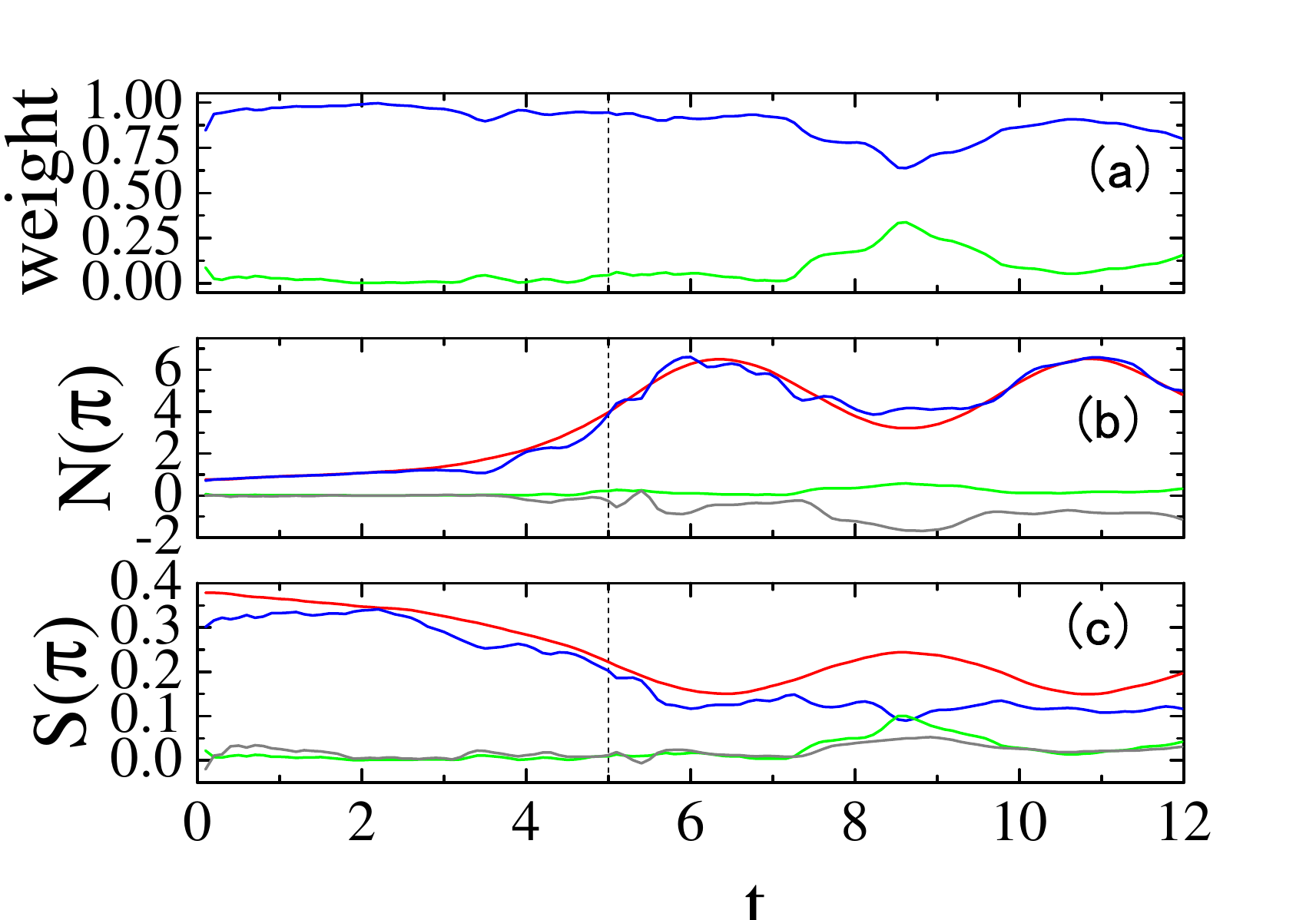}
\caption{(Color online) $N=14$, $\tau=5$: weights of the two main eigenmodes of the overlap matrix (a), and their contributions to $N(\pi)$ (b) and $S(\pi)$ (c). The red line is the total, blue and green are the diagonal contributions of modes 1 and 2, and the gray line is their interference.}
\label{fig:modes}
\end{figure}

The calculation for $N=14$ was carried out for $\tau=5$ up to $t=12$, that is, seven time units after the end of the ramp. Exact diagonalization is not available for this size. First, we examine the effect of the reconstruction (Sec.~\ref{sec:recon} and Appendix~\ref{app:recon}) by comparing with the original run continued without reconstruction (dotted lines in Fig.~\ref{fig:corr14}). The energies (Fig.~\ref{fig:corr14}(a)) differ by $1.4\times10^{-4}$ per site at $t=12$, and the correlation functions agree within about 1\% (rms) for $5\le t\le12$, so that the reconstruction does not change the state.

Next, we discuss the dynamics. The correlation functions (Fig.~\ref{fig:corr14}) show the same picture as for $N=10$. The blue dashed lines in Fig.~\ref{fig:corr14} are the static ResHF solutions at the instantaneous value $V(t)$ ($V=2.5$ after the ramp). During the first part of the ramp ($V\lesssim2.3$) the correlation functions follow the static values closely, but in the second half of the ramp they lag behind: at the end of the ramp, $N(\pi)=4.0$ against the static value $6.8$, and $S(\pi)=0.22$ against $0.14$. After the ramp, $N(\pi)$ oscillates between $3.2$ and $6.6$ and $S(\pi)$ between $0.15$ and $0.24$, so that the static values lie on the outer side of the oscillation; $B(\pi)$ oscillates between $3.9$ and $6.8$, well above the static value $2.6$. The energy after the ramp, $20.86$, exceeds the static energy at $V=2.5$, $20.52$, by $0.34$ (1.7\%). The dimerization amplitude (Fig.~\ref{fig:D14}) is $D=0.207$ at $t=0$ and $0.202$ at $t=12$, close to the initial value, whereas the static value at $V=2.5$ is $0.017$, about twelve times smaller than the dynamic $D$ at $t=12$. The lattice therefore stays almost frozen while the electrons respond, as for $N=10$.

After the ramp, $N(\pi)$ oscillates with maxima at $t\approx6.3$ and $10.9$ and a minimum at $t\approx8.6$, $S(\pi)$ oscillates in antiphase, and $B(\pi)$ follows. What changes during this oscillation can be seen in the configurations and their weights. Figure~\ref{fig:conf} shows the charge $n_\uparrow+n_\downarrow$ and the spin $n_\uparrow-n_\downarrow$ at each site, for each of the five configurations, at the maximum ($t=6.30$, a), at the minimum ($t=8.62$, b), and at the next maximum ($t=10.88$, c) of $N(\pi)$. Configurations 1, 2, and 5 have a strong charge-density wave, configuration 3 has a weak charge modulation and a spin density localized on a few sites, and configuration 4 lies in between. The shapes of the configurations are almost the same at the maxima and at the minimum. The oscillation of $N(\pi)$ and $S(\pi)$ is therefore not caused by a change of the configurations, and has to be found in their weights.

Because the configurations are not orthogonal, their weights have to be defined with care. We use the weights on the eigenmodes of the overlap matrix $S$, $w_k=\lambda_k|(V^\dagger C)_k|^2/\langle\Psi|\Psi\rangle$, where $C$ is the coefficient vector and $\lambda_k$ and the columns of $V$ are the eigenvalues and eigenvectors of $S$ for normalized configurations (L\"owdin-type weights~\cite{Lowdin}). These weights are non-negative and sum to one.

Two modes carry almost all of the weight (Fig.~\ref{fig:modes}). Mode 1 is essentially the combination of configurations 1, 2, and 5, and mode 2 is made of configurations 3 and 4. During the ramp the weight of mode 1 is 0.93--0.99. After the ramp the weight of mode 2 grows from 0.04 at the first maximum of $N(\pi)$ to 0.34 at the minimum, and the weight of mode 1 decreases from 0.92 to 0.64. The decrease of $N(\pi)$ at the minimum comes from the reduction of mode 1 and from the interference between the two modes (the contribution of mode 2 itself is positive, $+0.57$), whereas the increase of $S(\pi)$ comes almost entirely from mode 2 ($+0.088$ of the total $+0.093$; mode 1 gives $-0.035$ and the interference $+0.036$). The oscillation of the charge and spin correlations after the ramp can thus be described as the oscillating transfer of weight between the group of configurations with a strong charge-density wave and the group with a weak charge-density wave and localized spin density, while each configuration, including its lattice distortion, changes little in time. The two groups also differ in the bond order and in the lattice (Table~\ref{tab:confD}): at any given time, in configurations 3 and 4 the alternation of the bond order is larger than in configurations 1, 2, and 5 (about twice as large in configuration 3), and the lattice dimerization $D$ is also larger, by 10--15\%. The increase of $B(\pi)$ together with $S(\pi)$ at the minimum of $N(\pi)$ is therefore consistent with the gain of weight of mode 2. The second maximum ($t=10.88$) can be analyzed in the same way. That the oscillation can be read from the configurations and the coefficients in this way is a strength of the method (see Appendix~\ref{app:recon} for the non-uniqueness of the expansion).

\clearpage
\begin{table}[t]
\caption{$N=14$, $\tau=5$: lattice dimerization $D$ of each configuration and alternation of its bond order, $|N^{-1}\sum_\ell(-1)^\ell B_\ell|$, at the maximum ($t=6.30$), the minimum ($t=8.62$), and the next maximum ($t=10.88$) of $N(\pi)$. $D=2|Q_\pi|$ is evaluated from the coherent state of the configuration, and $B_\ell$ from its Slater determinant.}
\label{tab:confD}
\begin{ruledtabular}
\begin{tabular}{lcccccc}
 & \multicolumn{3}{c}{$D$} & \multicolumn{3}{c}{Bond-order alternation} \\
Configuration & 6.30 & 8.62 & 10.88 & 6.30 & 8.62 & 10.88 \\
\hline
1 & 0.198 & 0.189 & 0.195 & 0.270 & 0.324 & 0.245 \\
2 & 0.200 & 0.187 & 0.192 & 0.238 & 0.307 & 0.216 \\
3 & 0.228 & 0.216 & 0.221 & 0.537 & 0.560 & 0.467 \\
4 & 0.217 & 0.210 & 0.220 & 0.316 & 0.363 & 0.387 \\
5 & 0.200 & 0.192 & 0.193 & 0.267 & 0.343 & 0.217 \\
\end{tabular}
\end{ruledtabular}
\end{table}

\clearpage
\section{Discussion and Summary}\label{sec:discussion}

We have developed a time-dependent resonating Hartree--Fock (TD-ResHF) method, in which the wave function is a superposition of a small number of nonorthogonal configurations (five in this work), each the product of a Slater determinant and a coherent state of the lattice. The coefficients, the orbitals, and the coherent-state amplitudes are all determined by the McLachlan variational principle. The evolution is stable with a reconstruction of the configuration set, which is applied once or twice in a run.

We applied the method to the extended Hubbard model coupled to SSH-type phonons, in which the nearest-neighbor interaction $V$ is raised linearly in time. The method reproduces the exact time evolution for all ramp times: for $N=6$ the deviation of the correlation functions is below 1\% (all lattice modes), and for $N=10$ it is 1--2\% (reference with the $\pi$ phonon mode), including the oscillation after the ramp and the lattice distortion and energy. With the same number of configurations the method also treats $N=14$, a size too large for exact diagonalization, with stable evolution and a consistent physical picture.

The electrons and the lattice have different time scales. The correlation functions oscillate after the ramp with a period of about 4, which gives the time scale of the electrons, whereas the period of the $\pi$ phonon mode is $2\pi/\omega\approx95$. The ramps with $\tau=5$ and $13$ are much shorter than the phonon period. The lattice stays almost frozen during the ramp while the electrons change, and the correlation functions at the end of the ramp are those of the electrons moving in an almost frozen lattice, not those of the static state at the final $V$. The charge-density wave is still the dominant correlation, but it has grown less than in the static state, the spin correlation has decreased less, and the bond-order correlation, which is favored by the dimerized lattice, is enhanced. The electrons are excited by the ramp (strongly for $\tau=5$ and weakly for $\tau=13$), but not by the motion of the lattice.

For $\tau=100$, comparable to the phonon period, the lattice moves during the ramp, and the electrons follow both the ramp and the motion of the lattice adiabatically, whereas the lattice does not reach the static state: $D$ remains well above its static value (Fig.~\ref{fig:D}(c)). The electrons alone already follow the ramp adiabatically for $\tau=30$ (Table~\ref{tab:post}), so that the deviation from the static state is due to the lattice. The deviation consists of two parts. The oscillation after the ramp reflects the nonadiabaticity of the electrons and decreases with $\tau$, whereas the offset of the center of the oscillation from the static value is set by the difference between the remaining and the relaxed lattice distortion, and remains until the lattice follows the ramp, which has happened only partly for $\tau=100$. The deviation is small for $N=6$, where the Peierls-type transition is hardly visible, and larger for $N=10$ and $14$. These regimes are the familiar picture of fast electrons and a slow lattice (Born--Oppenheimer picture). TD-ResHF does not assume this separation, and finds it in the results.

The wave function itself can be examined: for $N=14$ the oscillation after the ramp is a transfer of weight between two groups of configurations, one with a strong charge-density wave and one with localized spin density, whereas the configurations hardly change. This kind of information is not available from the correlation functions alone, and is a strength of the method. The expansion of a state in nonorthogonal configurations is not unique, but the same wave function leads to the same physics for any such expansion (Appendix~\ref{app:recon}).

The static results also suggest a bond-order-wave ground state. For $N=10$, $B(\pi)$ increases with $V$, has a maximum at $V\approx2$, and then decreases, while $N(\pi)$ increases and $S(\pi)$ decreases monotonically. In the half-filled one-dimensional extended Hubbard model without phonons, a bond-order-wave (BOW) phase is found between the spin-density-wave and the charge-density-wave phases, near $V\approx U/2$, for intermediate values of $U$ \cite{Nakamura,Sengupta,Jeckelmann,SandvikBOW}. Since we use finite clusters, we cannot discuss the phase transitions. The rise and fall of $B(\pi)$ with $V$ suggests, however, that a BOW ground state exists in this range of $V$.

In the model without a heat bath, the excess energy cannot leave the system, and the lattice does not relax to the static state even after the ramp: for $N=14$ the dimerization amplitude is still about twelve times larger than the static value seven time units after the end of the ramp. This is a property of the coupled equations of motion. What our results show is that a ramp that is fast compared with the lattice period leads to a state that differs strongly from the static state of the final Hamiltonian, and that this state persists over the times studied here. We have not studied the dependence on the model parameters, the lattice frequency $\omega$ in particular. Whether the long-range order of the infinite system behaves in the same way is a subject of future work.

The configurations become nearly linearly dependent after the ramp, and in many cases the evolution is stable only with the reconstruction of Sec.~\ref{sec:recon} and Appendix~A. The reconstruction does not change the physical quantities (the energy with and without reconstruction differs by $1.4\times10^{-4}$ per site for $N=14$ at $t=12$), but it is an additional numerical step. The number of configurations, $\mathrm{NBM}=5$, was chosen by comparison with calculations with other numbers (Sec.~\ref{sec:nbm}). The method is not restricted to the present model: other electron--lattice couplings, more phonon modes, and larger systems can be treated in the same way. In this work we have presented the method and checked it.

\begin{acknowledgments}
The authors used the AI assistant Claude (Claude Sonnet 5.5, Anthropic) during this work. It wrote the exact-diagonalization code used for the comparisons and the figure data, and helped to draft and edit the manuscript text and its \LaTeX{} source. The time-evolution calculations were performed with the authors' own program. The authors verified the AI-assisted code and text and take full responsibility for the content of this paper.
\end{acknowledgments}
\appendix
\section{Reconstruction of the configuration set}
\label{app:recon}

The reconstruction is used only to keep the time evolution stable when the configurations become nearly linearly dependent. At a time $t_{\rm r}$ it replaces the set of configurations and coefficients by another set that represents the same state and is better conditioned, and the time evolution is restarted with the same equations of motion. This differs from approaches that discard the linearly dependent directions and evolve in a working space of varying dimension \cite{JoubertDoriol}, or that modify the equations of motion to restore the energy conservation \cite{Asaad} or to stabilize the evolution \cite{Hojlund}. This appendix describes the procedure.

The reconstruction is formulated as an optimization with four requirements. The new state agrees with the old state (the term res), and the expectation values of the observables agree as well (oerr). The energy is kept (the constraint on the coefficients). And the new configurations are kept apart from each other, so that the overlap matrix is well conditioned (pen and sep). The configurations and the coefficients are determined so that these requirements are satisfied together; the following subsections give the details.

\subsection{Setting}
At a time $t_{\rm r}$ the state is
\begin{equation}
|\Psi_{\rm old}\rangle=\sum_{m=1}^{M_c}C^{\rm old}_m|\Phi^{\rm old}_m\rangle,\qquad
|\Phi_m\rangle=\hat P\bigl[|U^\uparrow_m\rangle|U^\downarrow_m\rangle|z_m\rangle\bigr],
\end{equation}
where $U^\sigma_m$ is an $N\times N/2$ orbital matrix (Slater determinant), $|z_m\rangle$ is the lattice coherent state with the coordinates and momenta $(\bm q_m,\bm p_m)$, and $\hat P=N_G^{-1}\sum_g\hat g$ projects onto the symmetries of the Hamiltonian (translation, reflection, and spin flip). We define $S_{ij}=\langle\Phi_i|\Phi_j\rangle$, $H_{ij}=\langle\Phi_i|\hat H|\Phi_j\rangle$, and $O^{(k)}_{ij}=\langle\Phi_i|\hat O_k|\Phi_j\rangle$ for the observables $\hat O_k\in\{QQ_d,SS_d,n_{\uparrow\downarrow}\}$, with $QQ_d=N^{-1}\sum_in_in_{i+d}$, $SS_d=N^{-1}\sum_iS^z_iS^z_{i+d}$, and $n_{\uparrow\downarrow}=N^{-1}\sum_in_{i\uparrow}n_{i\downarrow}$ (double occupancy). Expectation values are $\langle O\rangle=C^\dagger OC/C^\dagger SC$, and $E_{\rm old}=C^{\rm old\dagger}H^{\rm old}C^{\rm old}/C^{\rm old\dagger}S^{\rm old}C^{\rm old}$.

\subsection{New configurations}\label{app:newconf}
Only the configurations in a free set $\mathcal F$ are replaced, and the others are kept. For $j\in\mathcal F$ we start from the old configuration $j$, $(U^{\sigma,0}_j,\bm q^0_j,\bm p^0_j)=(U^{\sigma,\rm old}_j,\bm q^{\rm old}_j,\bm p^{\rm old}_j)$, and define the new configuration with variables $x_j=(X^\uparrow_j,X^\downarrow_j,\delta\bm q_j,\delta\bm p_j)$ by
\begin{equation}
U^\sigma_j=\mathrm{orth}\bigl(U^{\sigma,0}_j+X^\sigma_j\bigr),\quad
\bm q_j=\bm q^0_j+\delta\bm q_j,\quad
\bm p_j=\bm p^0_j+\delta\bm p_j,
\end{equation}
where $\mathrm{orth}$ denotes the orthonormalization by QR decomposition.

\subsection{Coefficients for given configurations}
Let $|\Psi_{\rm new}\rangle=\sum_iC_i|\Phi^{\rm new}_i\rangle$ be the new state, with the coefficients $C$ to be determined, and let $S$ and $H$ be the matrices of the new configurations. With $B_{im}=\langle\Phi^{\rm new}_i|\Phi^{\rm old}_m\rangle$ and $b=BC^{\rm old}$, the normalized squared distance between the new and old states is
\begin{equation}
\mathrm{res}(C)=\frac{\|\Psi_{\rm new}-\Psi_{\rm old}\|^2}{\|\Psi_{\rm old}\|^2}
=1+\frac{C^\dagger SC-2\,\mathrm{Re}\,C^\dagger b}{S^{\rm old}_1},
\qquad S_1^{\rm old}=\|\Psi_{\rm old}\|^2 .
\end{equation}
The energy is kept by the constraint
\begin{equation}
g(C)\equiv C^\dagger\bigl(H-E_{\rm old}S\bigr)C=0 .
\end{equation}
Minimizing $\mathrm{res}$ under this constraint with a Lagrange multiplier $\lambda$ gives
\begin{equation}
C(\lambda)=\bigl[S+\lambda(H-E_{\rm old}S)\bigr]^{-1}b,
\end{equation}
where $\lambda$ is determined by the condition $g(C(\lambda))=0$, so that the energy of the new state is $E_{\rm old}$ ($\lambda=0$ is the unconstrained projection $C=S^{-1}b$). The solution of $g(C(\lambda))=0$ is found by a scan in $|\lambda|$ followed by Brent's method.

\subsection{Cost function}
The configurations are optimized by minimizing
\begin{equation}
J(x)=\mathrm{res}+10^{-2}\,\mathrm{pen}+W_{\rm obs}\,\mathrm{oerr}+W_{\rm sep}\,\mathrm{sep},
\end{equation}
\begin{align}
\mathrm{pen}&=\Bigl[\max\bigl(0,\log_{10}(\tau_0/\mu_0)\bigr)\Bigr]^2,\quad
\mu_0=\frac{\lambda_{\min}(S)}{\max_iS_{ii}},\\
\mathrm{oerr}&=\sum_d(\Delta QQ_d)^2+25\sum_d(\Delta SS_d)^2+25(\Delta n_{\uparrow\downarrow})^2,\quad
\Delta O=\langle O\rangle_{\rm new}-\langle O\rangle_{\rm old},\\
\mathrm{sep}&=\sum_{i<j}\max\bigl(0,G_{ij}-G_{\rm cap}\bigr)^2,\quad
G_{ij}=\frac{|S_{ij}|}{\sqrt{S_{ii}S_{jj}}}.
\end{align}
The term $\mathrm{pen}$ keeps the smallest eigenvalue $\lambda_{\min}(S)$ of the overlap matrix, normalized by the largest diagonal element, above $\tau_0$ (the penalty is proportional to the square of the number of orders of magnitude by which the eigenvalue falls below $\tau_0$), and $\mathrm{sep}$ keeps the normalized overlaps below $G_{\rm cap}$, that is, it keeps the configurations apart. We used $\tau_0=0.1$, $G_{\rm cap}=0.5$, and $W_{\rm sep}=1$ (in some cases a larger cap, see Table~\ref{tab:recon}). These values and the weights in $J$ were chosen empirically; changing the cap, the weights, and the set of configurations that are replaced gave the same time evolution as long as the residual res was small. $J$ is minimized by the L-BFGS-B method with finite-difference gradients (step $10^{-7}$).

\subsection{Final coefficients}\label{app:finalC}
With the optimized configurations fixed, the coefficients are recomputed from
\begin{equation}
\min_C\ \bigl[C^\dagger SC-2\,\mathrm{Re}\,C^\dagger b\bigr]\quad\text{subject to}\quad C^\dagger(H-E_{\rm old}S)C=0
\end{equation}
(sequential least-squares programming), followed by the normalization $C\to C/\sqrt{C^\dagger SC}$. The set $\{C,U^\sigma_j,\bm q_j,\bm p_j\}$ is the initial condition of the time evolution that is restarted at $t_{\rm r}$. No term of the equations of motion is modified; only the representation of the same state changes.

\subsection{Accuracy}
Because the coefficients are obtained by the projection above, the fidelity between the old and new states is, to first order in $\mathrm{res}$,
\begin{equation}
F=\frac{|\langle\Psi_{\rm old}|\Psi_{\rm new}\rangle|^2}{\|\Psi_{\rm old}\|^2\|\Psi_{\rm new}\|^2}\simeq1-\mathrm{res}.
\end{equation}
Table~\ref{tab:recon} lists the reconstructions used in this work. In all cases the energy is reproduced by the constraint (to $\sim10^{-10}$ where checked), $|\Delta QQ_d|\le4\times10^{-4}$, $|\Delta SS_d|\le2\times10^{-4}$, and $F\ge0.998$. The largest normalized overlap among the configurations is reduced to about 0.5 (0.73 in the case with a relaxed cap), and the smallest eigenvalue of the normalized overlap matrix is raised to about 0.1 for $N=10$ (0.18 for $N=14$). No reconstruction was used for $N=6$.

\begin{table}[t]
\caption{Reconstructions performed. $t_{\rm r}$ is the time of the reconstruction, res, $G_{\max}$ (largest normalized overlap), and $\mu_0$ (smallest eigenvalue of the normalized overlap matrix) are the final values of the optimization, and $\max|\Delta QQ|$ is the largest change of $QQ_d$.}
\label{tab:recon}
\begin{ruledtabular}
\begin{tabular}{llccccc}
System & $t_{\rm r}$ & res & $1-$res & $\max|\Delta QQ|$ & $G_{\max}$ & $\mu_0$ \\
\hline
$N{=}10$, $\tau{=}5$, NBM=5   & 5.0 (end of ramp) & $3.3\times10^{-4}$ & 0.9997 & $2\times10^{-4}$ & 0.5003 & 0.099 \\
$N{=}10$, $\tau{=}5$, NBM=5   & 14.0 & $1.3\times10^{-3}$ & 0.9987 & $2\times10^{-4}$ & 0.7325 & 0.091 \\
$N{=}10$, $\tau{=}13$, NBM=5  & 28 & $3.2\times10^{-4}$ & 0.9997 & $1\times10^{-4}$ & 0.5004 & 0.099 \\
$N{=}10$, $\tau{=}100$, NBM=5 & 2.5 & $1.2\times10^{-3}$ & 0.9988 & $2\times10^{-4}$ & 0.5002 & 0.095 \\
$N{=}10$, $\tau{=}100$, NBM=5 & 66.5 & $3.9\times10^{-4}$ & 0.9996 & $1\times10^{-4}$ & 0.5006 & 0.098 \\
$N{=}10$, $\tau{=}5$, NBM=7   & 0.25 & $7.3\times10^{-5}$ & 0.99993 & $<5\times10^{-5}$ & 0.5001 & 0.100 \\
$N{=}10$, $\tau{=}5$, NBM=7   & 5.5 & $5.1\times10^{-4}$ & 0.9995 & $4\times10^{-4}$ & 0.5001 & 0.098 \\
$N{=}14$, $\tau{=}5$, NBM=5   & 5.0 (end of ramp) & $3.5\times10^{-4}$ & 0.9997 & $1\times10^{-4}$ & 0.5008 & 0.180 \\
\end{tabular}
\end{ruledtabular}
\end{table}

\subsection{Configurations before and after the reconstruction}\label{app:conf}
\begin{figure}[t]
\centering
\includegraphics[width=0.85\columnwidth]{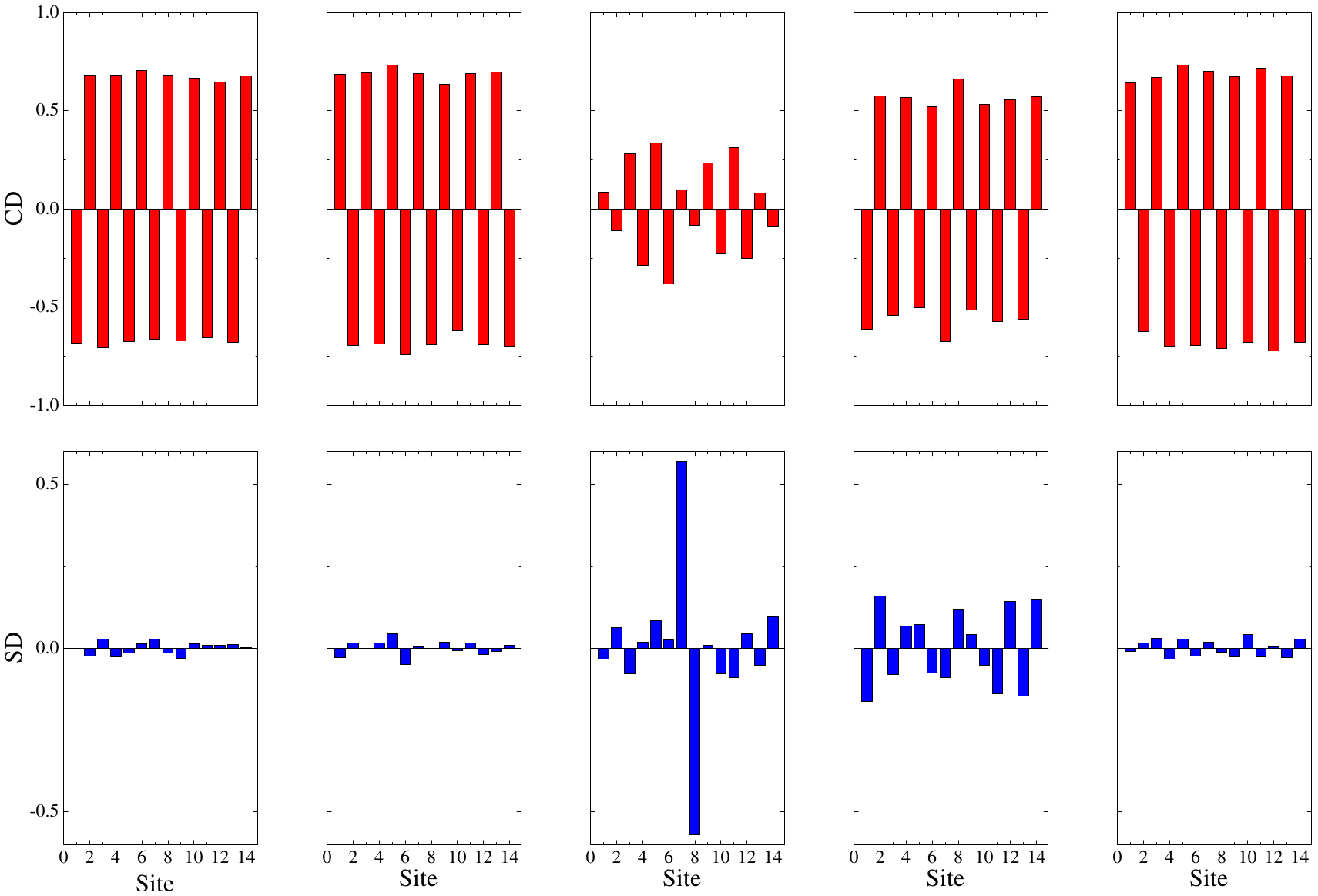}
\caption{(Color online) Configurations of the $N=14$ wave function at the second maximum of $N(\pi)$ ($t=10.88$) in the run without reconstruction. For each of the five configurations, the charge $n_\uparrow+n_\downarrow$ (CD, shown relative to 1) and the spin $n_\uparrow-n_\downarrow$ (SD, relative to 0) at each site. To be compared with Fig.~\ref{fig:conf}(c) (with the reconstruction at $t=5$).}
\label{fig:norecon}
\end{figure}

The reconstruction changes the configurations, whereas the wave function is the same. Figure~\ref{fig:norecon} shows the configurations of the $N=14$ wave function at the second maximum of $N(\pi)$ ($t=10.88$) in the run without reconstruction, and Fig.~\ref{fig:conf}(c) shows those of the run with the reconstruction at $t=5$ at the same time. The two sets are classified in the same way: configurations 1, 2, and 5 have a strong charge-density wave, configuration 3 has a weak charge modulation and a spin density localized on a few sites, and configuration 4 lies in between. The sites at which the spin is localized differ between the two sets, but the wave function is symmetrized with respect to the translations, so that the position of the localized spin has no physical meaning. The physical content that is read from the configurations is therefore the same with and without the reconstruction, as is the time evolution of the correlation functions and the energy (Fig.~\ref{fig:corr14}).

\subsection{Procedure}
The reconstruction at a time $t_{\rm r}$ proceeds as follows.
\begin{enumerate}
\item Compute $S^{\rm old}$, $H^{\rm old}$, and $O^{(k),\rm old}$ from the state at $t_{\rm r}$, and the quantities $S_1^{\rm old}$, $E_{\rm old}$, and $\langle O_k\rangle_{\rm old}$.
\item Set the optimization variables $x=0$. The variables $x$ are the displacements $x_j$ of the configurations $j\in\mathcal F$ from the old configurations, defined in Sec.~\ref{app:newconf}, so that $x=0$ is the old configuration itself.
\item Optimize the configurations by the L-BFGS-B method. One iteration, for given $x$, consists of: (a) build the configurations $\Phi_j(x)$ for $j\in\mathcal F$; (b) compute $S$, $H$, $O^{(k)}$, and $B$ (only the matrix elements that contain a changed configuration); (c) find $\lambda$ with $g(C(\lambda))=0$ and evaluate the coefficients $C(\lambda)$, which are used only to evaluate the cost function; (d) evaluate res, pen, oerr, sep, and $J$; (e) compute the gradient of $J$ by finite differences (parallel over the variables) and update $x$. The iteration is repeated until L-BFGS-B converges or the iteration limit is reached, and the result is the set of optimized configurations.
\item With the optimized configurations fixed, determine the final coefficients $C$ by the constrained minimization of Sec.~\ref{app:finalC} and normalize them.
\item Restart the time evolution at $t_{\rm r}$ from $\{C,\Phi_j\}$.
\end{enumerate}

\section{Exact diagonalization and its tests}
\label{app:ed}

\subsection{Method}
The reference solution used in Sec.~\ref{sec:results} is obtained by exact diagonalization (ED) of the same extended Hubbard--SSH model as in the main text, on a ring of $N$ sites with $N/2$ electrons of each spin. The lattice is expanded in the normal modes of the ring with nonzero frequency, $\omega_k=\omega\,|\sin(k/2)|$; the uniform translation does not couple to the electrons. For $N=6$ all five modes are kept. For $N=10$ the solution with all modes is not feasible, and only the zone-boundary ($\pi$) mode is kept. The error of this approximation is estimated for $N=6$ by comparison with the solution with all modes (Sec.~\ref{sec:valid}). The phonon space is truncated at $M_{\rm cut}$ phonons in total, and the lattice energy is measured without the zero-point energy. The ground state at $V=1.5$ is obtained by the Lanczos method, and the time evolution is calculated by the Chebyshev expansion of $e^{-iH\Delta t}$ with $\Delta t=0.05$, where $V$ is taken at the midpoint of each step. The ramp is $V(t)=V_i+(V_f-V_i)\,t/\tau$ with $V_i=1.5$ and $V_f=2.5$ (or $V_i=2.0$ for $\tau=100$).

The phonon cutoff $M_{\rm cut}$ was chosen as follows. For $N=6$ with all modes, $M_{\rm cut}=10$ gives $1.2\times10^6$ states; at $V=1.5$, $B(\pi)$ is $2.86634$ for $M_{\rm cut}=9$ and $2.86671$ for $M_{\rm cut}=10$, and the remaining error is estimated to be about $3\times10^{-4}$ (0.01\%) from the geometric decrease of the successive differences, while $N(\pi)$ and $S(\pi)$ change by less than $5\times10^{-6}$ and $4\times10^{-5}$. For $N=6$ with the $\pi$ mode only, the same $M_{\rm cut}=10$ is used. For $N=10$ (the $\pi$ mode only), we checked the convergence for $\tau=5$: at $t=12$ the correlation functions $N(\pi)$, $S(\pi)$, and $B(\pi)$ change by less than $10^{-3}$ between $M_{\rm cut}=24$ and $32$, and when $\Delta t$ is halved. The calculations for $N=10$ use $M_{\rm cut}=32$.

\subsection{Tests}
The ED code was written with the help of an AI assistant. Because the reference solution relies on it, it was checked against independent calculations. Table~\ref{tab:edtests} summarizes the tests.

\begin{table}[t]
\caption{Tests of the exact-diagonalization code. $N$ is the number of sites, $M_{\rm cut}$ the phonon cutoff.}
\label{tab:edtests}
\begin{ruledtabular}
\begin{tabular}{p{2.6cm}p{3cm}p{4.2cm}p{4.5cm}}
Test & System & Compared with & Result \\
\hline
1. Ground-state energy & $N=10$, $U=4$, $V=0$ & Lieb--Wu equations, finite ring & $-5.8343218$ vs $-5.8343226$ (difference $8.5\times10^{-7}$, constant in $U$)\\
2. Double occupancy & $N=10$, $U=4$, $V=0$ & Hellmann--Feynman, $\partial E/\partial U$ from the Lieb--Wu energy at $U=4\pm0.001$ & $1.0408490$ vs $1.0408491$ (sum over sites)\\
3. Lowest four levels & $N=6$, $U=4$, $V=2$ & Independent Fortran code (full Fock space) & agree to 10 digits\\
4. Lowest levels & $N=4$, $U=4$, $V=2$ & Independent Fortran code & agree to 10 digits (two doubly degenerate levels)\\
5. Electron + $\pi$ phonon & $N=4$ ($M_{\rm cut}=6$), $N=6$ ($M_{\rm cut}=4$), $U=4$, $V=2$ & Independent Fortran code, with phonons & lowest four levels agree to 10 digits\\
6. $\pi$ mode vs all modes & $N=6$, $U=4$, $V=2$ & ED with all phonon modes & $E$ differs by $4.6\times10^{-3}$, independent of $M_{\rm cut}=2$--$6$\\
\end{tabular}
\end{ruledtabular}
\end{table}

Tests~1 and 2 use the Bethe-ansatz (Lieb--Wu) solution of the one-dimensional Hubbard model, which is independent of any matrix construction. The Lieb--Wu equations~\cite{LiebWu} for the finite ring were solved numerically with a program originating from Ref.~\cite{Homopolar}; only energies are obtained from this solution, so the double occupancy was checked through the Hellmann--Feynman relation $\partial E/\partial U=\sum_l\langle n_{l\uparrow}n_{l\downarrow}\rangle$. The small constant energy difference in test~1 (relative $1.5\times10^{-7}$) does not depend on $U$ and is attributed to the tolerance of the root search. Tests~3--5 use a separately written Fortran program (full diagonalization with LAPACK) that was run by one of the authors. The coefficient of the electron--phonon coupling was checked through the static ResHF solution, whose dimerization amplitude agrees with the ED value (Sec.~\ref{sec:static}). Test~6 does not verify the code but gives the size of the $\pi$-mode approximation: the exact solution with all modes at $N=6$ differs from the $\pi$-only one by at most 2--7\% in the correlation functions (Sec.~\ref{sec:results}).

\subsection{Post-ramp oscillation due to electronic excitation}
\label{app:amp}

The oscillation after the ramp is a standard consequence of quantum mechanics: a state prepared by a ramp of finite duration contains a small admixture of excited states, and the interference between the ground state and the excited states oscillates at the excitation energy (see, e.g., Ref.~\cite{Polkovnikov}). We show this for the electrons alone ($N=10$) in order to confirm it quantitatively in the present model. After the ramp, $V=V_f=2.5$ is constant, and the state is expanded in the eigenstates $|n\rangle$ of the final Hamiltonian $H_f$ with energies $E_n$, $|\psi(t)\rangle=\sum_nc_ne^{-iE_nt}|n\rangle$, where the $c_n$ are fixed by the ramp. For an operator $\hat O$ with the matrix elements $O_{nm}$,
\begin{equation}
\langle\hat O\rangle(t)=\sum_n|c_n|^2O_{nn}+2\sum_{n<m}|c_nc_m|\,|O_{nm}|\cos\bigl[(E_m-E_n)t+\varphi_{nm}\bigr].
\label{eq:Ot}
\end{equation}
The first term is the mean value, and the second term, the interference between different eigenstates, is the oscillation. Its amplitude is proportional to the product of the probability amplitudes $c_n$ and $c_m$, and not to the product of the weights $|c_n|^2$ and $|c_m|^2$.

We expanded the exact state at the end of the ramp, $t=\tau$, in the 40 lowest eigenstates of $H_f$ (Lanczos method). Only the lowest excited state that couples to $N(\pi)$ has a sizable weight; it lies $\Omega=E_1-E_0=1.6938$ above the ground state, and $|N_{01}|=2.661$ for the operator $\hat N(\pi)$. The weights of all other excited states are below $10^{-4}$. The weight $p=|c_1|^2$ of this state is given in Table~\ref{tab:pweight}.
\begin{table}[t]
\caption{Weights of the ground state, $|c_0|^2$, and of the lowest excited state coupled to $N(\pi)$, $p=|c_1|^2$, in the exact state at the end of the ramp, and the excess energy $\Delta E=E-E_0$ divided by $\Omega$.}\label{tab:pweight}
\begin{tabular}{cccc}
\hline\hline
$\tau$ & $|c_0|^2$ & $p$ & $\Delta E/\Omega$ \\
\hline
5 & 0.9373 & 0.0623 & 0.0634\\
13 & 0.9965 & 0.00335 & 0.00360\\
30 & 0.9994 & 0.00057 & 0.00061\\
\hline\hline
\end{tabular}
\end{table}
Here $\Delta E=E-E_0$ is the excess energy. Eq.~\eqref{eq:Ot} with the two states gives, for $\hat O=\hat N(\pi)$,
\begin{equation}
\langle\hat N(\pi)\rangle(t)=(1-p)N_{00}+pN_{11}+2\sqrt{p(1-p)}\,|N_{01}|\cos(\Omega t+\varphi),
\end{equation}
so that the oscillation amplitude $A_N$ (half of the peak-to-peak value) is
\begin{equation}
A_N=2\sqrt{p(1-p)}\,|N_{01}|\simeq2\sqrt p\,|N_{01}|.
\end{equation}
The oscillation is proportional to $\sqrt p$, whereas the shift of the mean value is proportional to $p$; for $p\ll1$ a weight of a few tenths of a percent therefore already gives a visible oscillation.

The weight is related to the excess energy, which is conserved after the ramp: $\Delta E=\sum_n|c_n|^2(E_n-E_0)\simeq p\,\Omega$, that is, $p\simeq\Delta E/\Omega$ (last column of Table~\ref{tab:pweight}). Using $p$ of Table~\ref{tab:pweight}, $A_N\simeq2\sqrt p\,|N_{01}|=1.33$, $0.31$, and $0.13$ for $\tau=5$, 13, and 30, which agree with the amplitudes of the exact dynamics, $1.31$, $0.31$, and $0.13$ ($A_N$ in Table~\ref{tab:post}), to within 2\%. The angular frequency of the oscillation in the exact dynamics, 1.70--1.73 (period 3.6--3.7), is close to $\Omega$. Since the amplitude is determined by the weight $p$ of the excited state, which is related to the excess energy, the amplitude of the oscillation can serve as a measure of whether the electrons follow the ramp adiabatically.

\subsection{Fidelity of the TD-ResHF state with the exact solution}
\label{app:overlap}

For $N=10$ electrons the TD-ResHF state can be written in the same Fock space as the exact state and compared with it. Each configuration is a pair of Slater determinants (up and down spins). The amplitude of a basis state of the Fock space, specified by the occupied sites of each spin, is the product of the two $5\times5$ determinants of the corresponding rows of the orbital matrices. The amplitudes of the five configurations, with the translations, the reflection, and the spin flip applied (40 group elements in total), are summed with the coefficients $C_k$. The vector obtained in this way reproduces the energy and the correlation functions calculated from the TD-ResHF configurations and coefficients (the energy to within $10^{-6}$). The vectors of the TD-ResHF states at the end of the ramp, $t=\tau$, were compared with the exact state at the same time and expanded in the eigenstates of $H_f$ (the 72 lowest levels, with excitation energies up to about 4.8, which contain 99.99\% of the exact state). The vector of the initial state, the static solution at $V=1.5$, was expanded in the same way in the eigenstates of the Hamiltonian at $V=1.5$.

\begin{table}[t]
\caption{Fidelity of the TD-ResHF state with the exact state at the end of the ramp ($N=10$, electrons only) and weights in the eigenstates of $H_f$ (of $H$ at $V=1.5$ for the initial state). $F$ is the fidelity, $|c_0|^2$ the weight of the ground state, $p$ the weight of the lowest excited state that couples to $N(\pi)$, and $w_>$ the total weight of the states above $E-E_0=4.8$; ex and TD denote the exact and the TD-ResHF states.}\label{tab:overlap}
\small
\begin{tabular}{ccccc}
\hline\hline
state & $F$ & $|c_0|^2$ & $p$ & $w_{>}$\\
 & & (ex / TD) & (ex / TD) & (ex / TD)\\
\hline
initial ($V=1.5$) & 0.998 & 1 / 0.998 & --- & 0 / $1.4\times10^{-3}$\\
$\tau=5$ & 0.999 & 0.937 / 0.938 & 0.0624 / 0.0603 & $1\times10^{-4}$ / $1.1\times10^{-3}$\\
$\tau=13$ & 0.999 & 0.9965 / 0.9955 & 0.00335 / 0.00328 & $1\times10^{-5}$ / $1.1\times10^{-3}$\\
$\tau=30$ & 0.999 & 0.9994 / 0.9983 & 0.00057 / 0.00049 & $3\times10^{-6}$ / $1.1\times10^{-3}$\\
\hline\hline
\end{tabular}
\end{table}
In Table~\ref{tab:overlap}, $F=|\langle\Psi_{\rm ex}|\Psi_{\rm TD}\rangle|^2$, $p$ is the weight of the lowest excited state that couples to $N(\pi)$ (Sec.~\ref{app:amp}), and $w_>$ is the total weight of the states above $E-E_0=4.8$. The fidelity with the exact solution is 0.998 or higher in all cases, which is consistent with the accuracy of the correlation functions. The TD-ResHF state contains the excited state responsible for the oscillation with nearly the same weight as the exact state (within 3\% for $\tau=5$ and 13; for $\tau=30$ both weights are small, $5.7\times10^{-4}$ and $4.9\times10^{-4}$), so that the amplitude of the oscillation after the ramp is reproduced. The deviation of the fidelity $F$ from unity is mainly due to a weight of about $10^{-3}$ at energies above 4.8, which is already present in the initial state ($1.4\times10^{-3}$) and is therefore not generated in the time evolution.

\end{document}